\documentclass[twocolumn]{aastex62}

\usepackage{amsmath}
\usepackage{mathrsfs}
\usepackage{wasysym}
\usepackage{hyperref}
\usepackage{bm}

\newcommand{\refeq}[1]{Eq.~(\ref{eq:#1})}          
          
\newcommand{\reffig}[1]{Figure~\ref{fig:#1}}          
\newcommand{\refsec}[1]{Section~\ref{sec:#1}}
\newcommand{\refapp}[1]{Appendix~\ref{app:#1}}

\newcommand{\Te}{T_{\rm e}}
\newcommand{\xe}{x_{\rm e}}
\newcommand{\nele}{n_{\rm e}}
\newcommand{\NH}{N_{\rm H}}
\newcommand{\nH}{n_{\rm H}}
\newcommand{\nHI}{n_{\rm HI}}

\newcommand{\Lya}{{{\rm Ly}\alpha}}

\newcommand{\kB}{k_{\rm B}}

\def\rmd{\mathrm{d}}

\begin{document}

\title{Partially Ionized Gas at Equilibrium Powered by Ultraviolet Irradiation}

\author[0000-0003-2091-8946]{Liang Dai}
\affiliation{Department of Physics, University of California, 366 Physics North MC 7300, Berkeley, CA 94720, USA}

\correspondingauthor{Liang Dai}
\email{liangdai@berkeley.edu}
 
\begin{abstract}

We study the formation of a partially ionized zone behind the usual H II zone in photo-ionized gas for gas density intermediate between ordinary H II regions and AGN BLRs and without X-rays. While Lyman continuum photons are depleted before reaching this zone, Balmer continuum photons can ionize the $n=2$ hydrogen atoms as those are pumped by scattering-trapped Ly$\alpha$ photons. Ly$\alpha$ photons need to be replenished in a steady state, but fast radiative cooling at high gas temperature makes replenishment through electron collisional excitation of $n=2$ inefficient. Independently of the temperature, Raman scattering of incident continuum photons on the Lyman-series damping wings injects Ly$\alpha$ photons in the line core and pumps the $n=2$ state, which helps realize a partially ionized equilibrium at $5000$--$7000\,$K. Since the ratio between the incident radiation flux and gas density is limited by the dynamic effects of radiation pressure, significant ionization fraction is only realized at densities $\nH \gtrsim 10^7\,{\rm cm}^{-3}$. A large H$\alpha$/H$\beta$ line ratio results from such partially ionized gas as large Balmer line optical depths enhance the $n=3$ population relative to higher $n$ states. Partial ionization also provides the ideal condition for internally forming an intense and very broad Ly$\alpha$ line, which can power strong fluorescent metal emission lines observed from the Weigelt blobs of $\eta$ Carinae in the Galaxy, and from a candidate young super star cluster in the Cosmic Noon Sunburst galaxy. This phenomenon may also have implications for understanding the spectra of high-$z$ galaxies that show unusually dense ionized gas.

\end{abstract}


\section{Introduction}
\label{sec:intro}

At typical densities of the interstellar medium (ISM), gas can stay photo-ionized by Lyman continuum (LyC) photons emitted by hot stars. The spatial structure of photo-ionized gas is a well-studied subject in ISM physics, dating back to the work of \cite{Stromgren1939}. In an ionization bounded situation, a (nearly) fully ionized H II zone forms, shielding behind it a (nearly) fully neutral H I zone from the LyC photons. The thickness of the H II/H I transition is set by the mean free path of the LyC photons, and is geometrically thin for typical values of the ionization parameter $\mathcal{U}=\Phi_{\rm LyC}/(\nH\,c)$. 

\begin{figure}[ht!]
    \centering
    \includegraphics[scale=0.28]{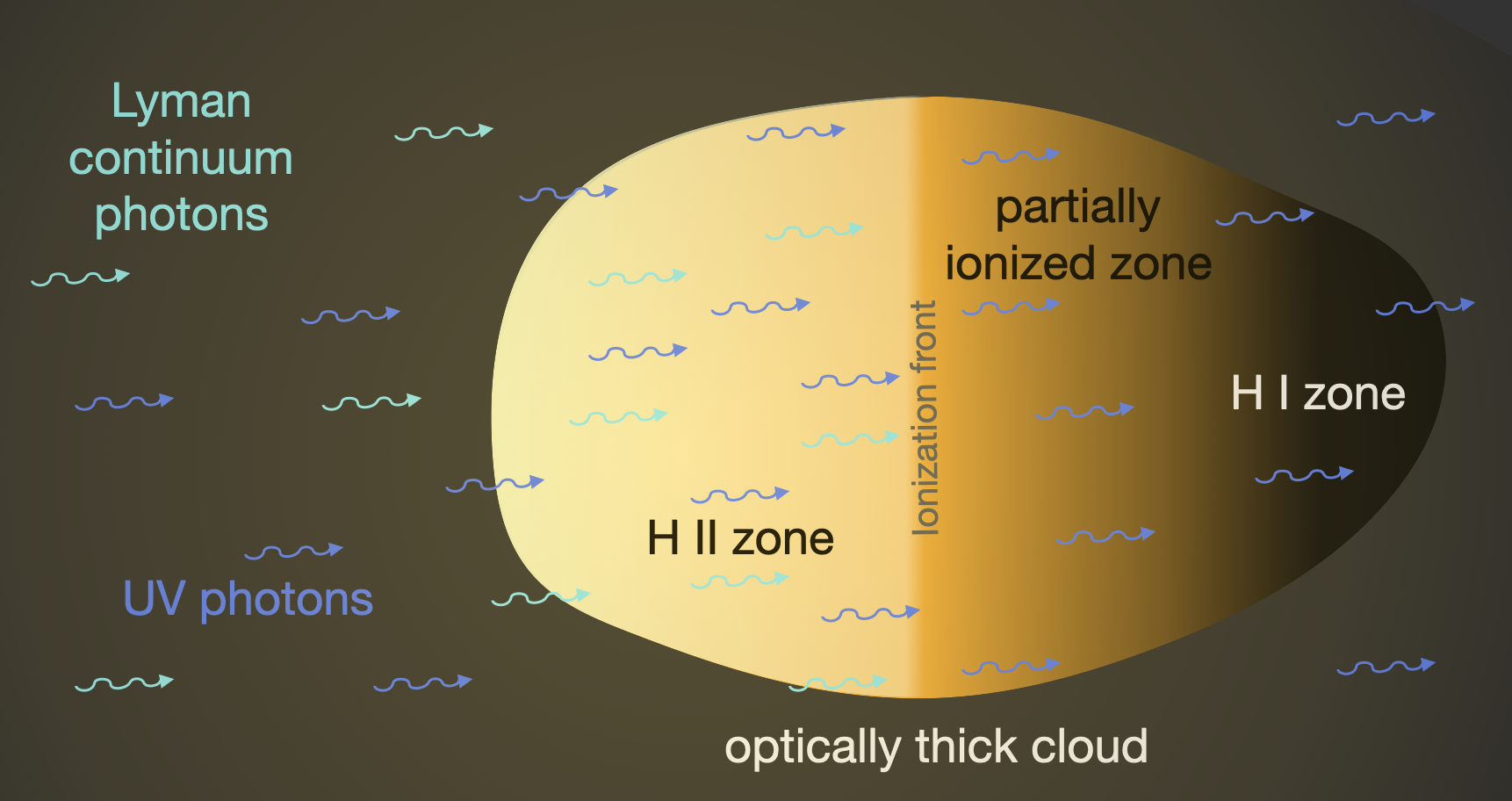}
    \caption{Ionization structure of a cloud optically thick to Lyman continuum photons but optically thin to UV photons.}
    \label{fig:cartoon}
\end{figure}

However, interstellar gas may be partially (not slightly) ionized under other conditions. A gas in collisional ionization equilibrium is partially ionized at intermediate gas temperatures $\Te=10000$--$20000\,$K, depending on the density, but without heating this state is not maintained for longer than the radiative cooling time. In this paper, we {\bf dissect} how photo-ionization can sustain a partially ionized zone of significant hydrogen column $\gtrsim 10^{20}\,{\rm cm}^{-2}$ behind the usual H II zone (shown in \reffig{cartoon}), which is not realized in ordinary Galactic H II regions.

Partial ionization behind fully ionized H II zone is known to arise in theoretical photo-ionization calculations of the Broad Line Regions (BLRs) in Active Galactic Nuclei (AGNs)~\cite[e.g. Figure 1 of][]{Ferland2003SpectrumOfSingleCloud}, which involve extremely dense gas with $n_{\rm H}\gtrsim 10^9$--$10^{10}\,{\rm cm}^{-3}$~\citep{OsterbrockFerland2006text}. It is widely recognized that the nebular physics there is complicated because of efficient collisional couplings which greatly enhance photo-ionization from excited H levels. Moreover, BLR clouds are irradiated by copious X-ray photons that penetrate into H I gas and broaden the H ionization front.

Here we study stellar irradiation conditions where penetrating X-rays are insignificant. Several observed nebular phenomena show evidence for the formation of a partially ionized zone even in such a case, which often involve a gas denser than in typical Galactic and extragalactic star-forming regions $\nH=10$--$10^4\,{\rm cm}^{-3}$~\citep{Kewley2013CosmicEvolutionOpticalLines}, but not quite as high as in AGN BLRs $n_{\rm H}\gtrsim 10^9\,{\rm cm}^{-3}$. One famous Galactic example is the Weigelt circumstellar gas blobs $300$--$1000\,$AU from the Luminous Blue Variable (LBV) $\eta$ Carinae~\citep{WeigeltEbersberger1986, Smith2004InnerHomunculus}, which have electron densities $\nele=10^7$--$10^8\,{\rm cm}^{-3}$ as inferred from comprehensive line ratio analyses~\citep{Hamann1999HSTSTISspectraEtaCarEjecta, Wallerstein2001LineIdenEtaCar, Hamann2012EtaCarInnerEjecta}. 

\cite{Verner2002FePhotoionModel} performed 1D photo-ionization modeling and analyzed the Fe II emission of the Weigelt blobs. They favored a model with a somewhat lower $\nele=10^6\,{\rm cm}^{-3}$, found a more smooth H II/H I transition in their photo-ionization model which enhances Fe II emission, but did not investigate the underlying cause. Strong fluorescent lines of both Fe II and Fe III are observed from the Weigelt blobs~\citep{Zethson2012Weigeltlines}. The Fe II emitting gas appears to be warm $\Te=6000$--$8000\,$K~\citep{Hamann2012EtaCarInnerEjecta}. \cite{Johansson2001FeIIpumping} and \cite{Klimov2002FeIIpumping} studied Fe II fluorescent lines pumped by Ly$\alpha$, and suggested that efficient pumping results from a scattering broadened Ly$\alpha$ line (half width $\gtrsim 500$--$600\,{\rm km}\,{\rm s}^{-1}$) that forms in a partially ionized zone between the H II and H I zones. Indeed, behind the usually sharp H II/H I front, the H I gas has a huge scattering optical depth for Ly$\alpha$ but emissivity is too low due to small ionization fraction and low gas temperature. Fe III fluorescent emissions pumped by Ly$\alpha$ also require significant broadening of the Ly$\alpha$ line (half width $\gtrsim 270\,{\rm km}\,{\rm s}^{-1}$) inside a zone of partial ionization to overcome detuning~\citep{Johansson2000FeIIIforbiddenlines}. However, those studies did not explain how the partially ionized zone forms and whether it can be much thicker than the LyC mean free path.

A recent discovery is the highly magnified source ``Godzilla''~\citep{Vanzella2020IonizeIGM, Diego2022godzilla, Sharon2022SunburstLensModel} in the Sunburst galaxy at $z=2.369$~\citep{Dahle2016}. Godzilla shows in its FUV spectrum strong fluorescent Fe III emission due to Ly$\alpha$-pumping~\citep{Vanzella2020Tr}, which hints at a photo-ionization condition similar to that of the Weigelt blobs. The physical nature of Godzilla is under active investigation: \cite{PascaleDai2024godzilla} suggest that the observed nebular emission is powered by a young massive star cluster magnified by about $500$--$2000$ fold, and find $\nele=10^7$--$10^8\,{\rm cm}^{-3}$ by analyziing metal forbidden line ratios. \cite{Choe2024} propose a post-eruption LBV star with circumstellar ejecta and subject to an extremely large magnification factor $\mu\sim 10^4$. Fluorescent O I lines as a result of Ly$\beta$ pumping are also observed from Godzilla~\citep{Diego2022godzilla, PascaleDai2024godzilla, Choe2024}. To allow efficient Ly$\alpha$ pumping, it is plausible that partial ionization also occurs in the nebula of Godzilla.

It is important to understand such spectral phenomena in order to correctly deduce the gas and radiation properties involved in those systems. This work aims to shed light on the mechanism of steady-state partial ionization under photo-ionization. The key insight is that beyond where incident LyC photons are depleted, hydrogen can stay partially ionized if the $n=2$ excited state of hydrogen is significantly populated and photo-ionized by UV light. A high $n=2$ population needs to be maintained by a high intensity of internal Ly$\alpha$ radiation, which, in the absence of photo-ionization of ground-state atoms, can be produced by electron collisional excitation of $n=2$ and by Raman scattering of incident FUV continuum photons on the damping wings of Ly$\beta$, Ly$\gamma$, Ly$\delta$, $\cdots$. Moreover, a steady state is possible only if energy injection by external radiation balances radiative cooling.  We will show that a sufficiently high gas density is necessary for significant ionization at equilibrium, if external radiation scales in proportion to the gas density. In particular, we highlight the unique role of Raman scattering when gas density is lower than in AGN BLR clouds ($\nH=10^7$--$10^9\,{\rm cm}^{-3}$) so that the populations of H I excited levels are not quite driven toward collisional equilibrium, and when penetrating X-ray photons are unavailable.

The remainder of this paper is organized as the following. In \refsec{one_zone}, we implement a one-zone model of hydrogen slab, investigate the condition for an equilibrium of partial ionization, and present analytic and numerical results. In particular, we will consider physical conditions relevant for $\eta$ Car's Weigelt blobs. In \refsec{cloudy}, we present \texttt{CLOUDY} calculations to verify simple one-zone calculations of the hydrogen slab, for parameters appropriate for the Weigelt blobs. We discuss our results in \refsec{discuss}, before providing concluding remarks in \refsec{concl}. Technical details are presented in Appendices for interested readers.

\section{One-Zone Model}
\label{sec:one_zone}

In reality, the partially ionized zone will develop profiles of gas temperature, ionization fraction and level populations that vary with the depth from the illuminated cloud surface. This can be treated by public photo-ionization codes such as \texttt{CLOUDY} which self-consistently computes ionization, level populations and temperature as a function of depth. To develop some simple understanding, we introduce in this Section a simplified model in which the partially ionized gas slab is assumed to have uniform physical and chemical properties.

Consider a slab with a hydrogen number density $\nH$, electron temperature $\Te$, and a thickness $d$. The slab is irradiated by continuum radiation with a spectrum $F_\nu(\nu)$ [${\rm erg}\,{\rm s}^{-1}\,{\rm cm}^{-2}\,{\rm Hz}^{-1}$]. Let us assume that no Lyman continuum photons with $h\nu > 13.6\,$eV reach the slab because they are completely absorbed in the usual H II zone that is expected to form in front of the partially ionized gas. However, continuum photons with $h\nu<13.6\,$eV may photo-ionize excited states of hydrogen. We shall assume that the slab is optically thin to photons in the Balmer continuum, Paschen continuum, and so on, which will be validated by our solutions.

\subsection{Hydrogen Atom Model and Level Populations}
\label{sec:levelpop}

We calculate for a small model of hydrogen atom that includes orbitals up to $n=5$. Since we mainly concern quite high densities $\nH \gtrsim 10^7\,{\rm cm}^{-3}$, we further assume that for the same $n$ the $l=0,1,2,\cdots,n-1$ levels are tightly coupled to each other through proton and electron impact and have populations proportional to their statistical weights $g_l=2\,(2\,l+1)$.

We solve for steady-state populations of the hydrogen states, as well as the ionization fraction $\xe$. The calculation takes into account photo-ionization, radiative recombination, collisional (de-)excitation and ionization by electron impact, and spontaneous and stimulated radiative transitions. Underlying equations are presented in \refapp{equil}.

We use cross sections tabulated in \cite{Burgess1965TableHydrogenicPxsec} for photo-ionization of ground and excited states of H and average over the $l$-levels by the statistical weights. For radiative recombination, we use the partial rate coefficients from \cite{Boardman1964RadiativeRecombinationHydrogen} for recombining into the ground and excited states. Since the slab must be optically thick to LyC photons (Case B), we ignore direct recombination to the ground state. For collisional (de-)excitation by electron impact, we use the effective collision strength data in \cite{Anderson2000ElectronImpactDataHI} and average over the $l$-levels. For collisional ionization from the various states, we implement the fitting formulae of \cite{Beigman1996ElectronCollisionIonization}.

\subsection{Resonant Line Transfer}
\label{sec:reslines}

We account for allowed radiative transitions between states. Depending on level populations and the slab column density $\NH=\nH\,d$, resonant lines may or may not be optically thick: all Lyman series lines are optically thick; if the $n=2$ population is significant, Balmer series may be optically thick too. We therefore adopt the Escape Probability Approximation (EPA) to model the trapping of all resonant lines~\citep{Rybicki1984book, ElitzurNetzer1985LineFluorescence, LockettElizur1989EscapeProbabilityFormalism}. In this treatment, the Einstein coefficient for spontaneous radiative transition is effectively reduced by a factor of the escape probability for a line photon to escape the gas slab. In this work, we prescribe a single-flight line escape probability~\citep{Kastner1993PhotonEscapeProbability}
\begin{align}
    \beta(\tau_0) = \begin{cases}
        1/(1+2\,\tau_0), & {\rm Case\,\,1}, \\
        1/(1+2\,\tau_0) + 0.25\,(a_v/\tau_0)^{1/2}, & {\rm Case\,\,2}.
    \end{cases}
\end{align}
Here $\tau_0$ is the optical depth at the resonant line center, and $a_v$ is the Voigt parameter. Case 1 and Case 2 apply to incomplete and complete frequency redistribution, respectively~\citep{ReesNetzerFerland1989BroadLineRegionAGN}. 

The EPA treatment is especially important for the Ly$\alpha$ line, the trapping of which enhances the $n=2$ population. We separately account for two-photon decay of the $2s$ state, which has a rate $A_{2s}=8.229\,{\rm s}^{-1}$. The resultant two-photon continuum is an important channel of cooling.

Internal absorption of line photons by dust grains might also be important. We account for this by prescribing an effective line escape probability~\citep{ReesNetzerFerland1989BroadLineRegionAGN}
\begin{align}
\label{eq:EPA_beta_eff}
    \beta^{\rm eff}(\tau_0) = \varepsilon + (1 - \varepsilon)\,\beta(\tau_0),
\end{align}
where $0\leqslant \varepsilon < 1$ is the probability of dust absorption during a single flight before the next scattering and is proportional to the wavelength-dependent cross section of dust absorption.

\subsection{Energy Equilibrium}
\label{sec:hcbalance}

Warm ionized gas continuously loses energy through radiation. Major radiative cooling channels relevant for the hydrogen gas are free-free emission, recombination continuum emission, two-photon continuum emission, and escape of resonant line photons. 

A steady state is possible only if energy is continuously provided to the system to compensate for radiative loss. For the case considered here, energy injection contributions include photon absorption by photo-ionization and by the inelastic Raman scattering. The latter process will be discussed in \refsec{Raman}.

For a steady state, requiring the balance between external energy injection and radiative loss is equivalent to requiring the balance of gas heating and cooling, which is realized directly through photo- and collisional ionization, recombination, and electron collisional (de-)excitation. In \refapp{energy}, we provide the equations for calculating the various energy injection and radiative loss contributions.

We have only considered processes involving the hydrogen atom. In a real gas, helium and metals make additional contributions to radiative cooling. For a gas slab shielded from all photons with $h\nu>13.6\,$eV, He is neutral. Since excited states of He are high-lying, radiative cooling through He is inefficient. In ordinary H II regions, metal line cooling is a dominant cooling mechanism, through optical forbidden lines of N II, O II, O III, S II, S III, etc. However, these forbidden lines are collisionally suppressed in the high density regime $\nele\gtrsim 10^6\,{\rm cm}^{-3}$ we concern. At lukewarm temperatures $\Te=5000$--$6000\,$K appropriate for the partially ionized gas, semi-forbidden and resonant lines in the FUV band are also suppressed. In \refsec{cloudy}, we present \texttt{CLOUDY} models including He and metals in the context of the Weigelt blobs in $\eta$ Car, which exhibit the phenomenon of partial ionization.

\subsection{Raman Scattering and Injected Ly$\alpha$ Photons}
\label{sec:Raman}

\begin{figure*}[ht!]
    \centering
    \includegraphics[width=\textwidth]{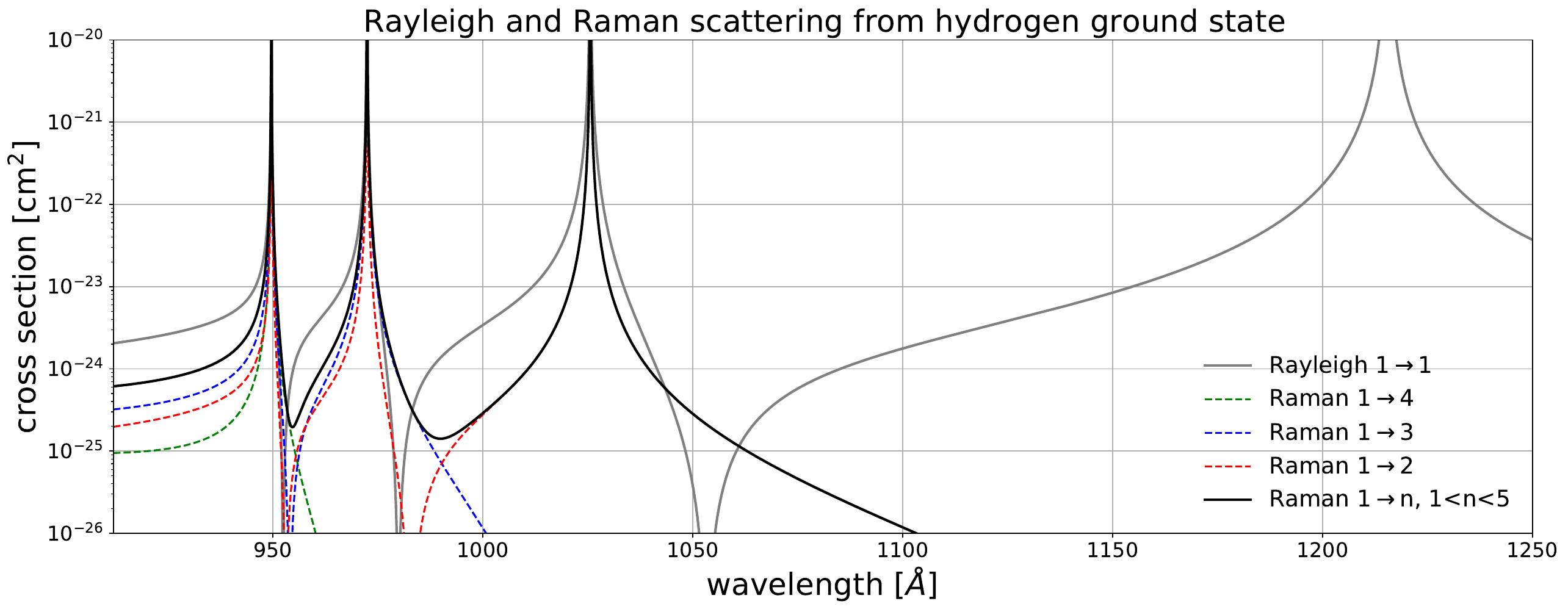}
    \caption{Cross sections of Rayleigh and Raman scattering channels from the $1s$ state of the hydrogen atom. Additional resonance peaks at the wavelengths of higher-order Lyman-series lines beyond Ly$\delta$ arise if more excited states are included in the hydrogen atom model. See \cite{Kokubo2024RayleighRaman} for cross sections computed using a larger atom model.}
    \label{fig:Rayleigh_Raman_xsec}
\end{figure*}

The slab typically has a large H I column density $N_{\rm HI} = (1-\xe)\,\nH\,d$, nearly all of which are in the ground state. Hydrogen atoms in the ground state do not absorb incident continuum photons below the Lyman limit $h\nu<13.6\,$eV through photo-ionization. However, at high $\nHI$, continuum photons on the damping wings of the Lyman lines can have substantial optical depths for both elastic (Rayleigh) and inelastic (Raman) scatterings.

Despite having a smaller cross section compared to Rayleigh scattering, Raman scattering contributes to net transitions from the ground state to the various excited states~\citep{Nussbaumer1989RamanScattering}. For example, an incident photon close to the Ly$\beta$ line may excite a $1s$ atom to the $2s$ state, which results in the emission of another photon close to H$\alpha$. This process is resonantly enhanced by a ``virtual'' $3p$ state, $1s\rightarrow 3p^*\rightarrow 2s$. There are additional resonantly enhanced Raman processes: $1s\rightarrow np^*\rightarrow 3s/3d$ for $n\geqslant 4$, $1s\rightarrow np^*\rightarrow 4s/4d$ for $n\geqslant 5$, and so on. Provided that the slab is optically thin to all Balmer continuum and higher-order continuum photons but efficiently traps all Lyman line photons, Raman processes result in a net ``injection'' of Ly$\alpha$ photons in the line core. Raman scattering has been detected from very broad wings of H$\alpha$ emission in H II regions in the Milky Way and in the Magellenic Clouds~\citep{Dopita2016RamanScattering, Henney2021RamanMapping}. Raman processes are also found important for the cosmological 21cm signal at Cosmic Dawn~\citep{Chen2004Lya21cm} and in the spectroscopic study of symbiotic stars (e.g.~\cite{Harries1996RamanSymbioticStars, Schmid1994RamanSymbioticStars, Lee2000RamanHaWingsSymbioticStars}). See \cite{Kokubo2024RayleighRaman} for a recent detailed discussion of the underlying atomic physics.

We adopt an analytic model to account for the effect of $N_{\rm HI}$ on incident photons. We use the analytic formulae in \cite{Nussbaumer1989RamanScattering} to calculate Rayleigh and Raman scattering cross sections (\reffig{Rayleigh_Raman_xsec}), including resonances up to $n=5$ to be consistent with the small hydrogen atom model used here but neglecting continuum virtual states~\citep{Nussbaumer1989RamanScattering}. With a large $N_{\rm HI}$, the slab can be optically thick to continuum photons in some wavelength ranges. Let $\tau_{\rm RR}(\nu)$ be the total optical depth of Rayleigh and Raman scattering processes from the $1s$ state, and $\epsilon_{\rm Ra}(\nu) < 1$ is the fractional contribution to $\tau_{\rm RR}(\nu)$ from all Raman processes. In the two-stream approximation of radiation transfer through a slab, the fraction of incident FUV photons of frequency $\nu$ that are Raman-absorbed in the gas and contribute to Ly$\alpha$ injection is
\begin{align}
    f_{\rm Ra}(\nu) = \frac{2}{ 1 + \epsilon_{\rm Ra}^{-1/2}(\nu)\,\coth\left(\frac12\,\epsilon_{\rm Ra}^{1/2}(\nu)\,\tau_{\rm RR}(\nu)\right) }.
\end{align}
The volume averaged Ly$\alpha$ injection rate from Raman processes $1s\rightarrow n'p^* \rightarrow ns/nd$ for a given $n\geqslant 2$ and all $n' > n$ is
\begin{align}
    A^{\rm Ra}_n = \frac{1}{(1-\xe)\,x_1\,\nH\,d}\,\int\rmd\nu\,\frac{F_\nu(\nu)}{h\nu}\,\frac{\sigma^{(n)}_{\rm Ra}(\nu)}{\sigma_{\rm Ra}(\nu)}\,f_{\rm Ra}(\nu).
\end{align}
where $x_1$ is the fractional population of the ground-state hydrogen, $F_\nu(\nu)$ is the incident flux and $\sigma^{(n)}_{\rm Ra}(\nu)$ is the cross section for Raman scattering from $1s$ as the initial state into the $n$-th state as the final state. The rate of net energy gain from all Raman processes, averaged over the slab volume, can be calculated using \refeq{GaRa}. In reality, the transfer and absorption of continuum photons in the gas slab varies with the depth, so our one-zone model is meant to be a crude treatment.

As we will see, Ly$\alpha$ injection by Raman scattering crucially helps to realize a steady state of partial ionization at low $\Te$. The gas still loses Ly$\alpha$ photons continuously due to photon escaping and internal destruction, and hence replenishment is required. Without photo-ionization by LyC photons, collisional excitation or collisional ionization are needed for that, which however is efficient only at high $\Te$. Thus, maintaining an equilibrium at a high $\Te$ may not be possible as gas heating can fall short of radiative cooling. Raman scattering replenishes Ly$\alpha$ photons regardless of the gas temperature, allowing energy balance to be satisfied.

\subsection{Low Temperature Limit}
\label{sec:lowTe}

If the equilibrium temperature $\Te$ is very low, electron collisions do not significantly ionize ground-state hydrogen atoms or excite them to excited states. As shown in \refapp{LowTe}, when all collisional processes are neglected the ionization fraction is given by
\begin{align}
\label{eq:xe_lowTe}
    & \frac{\xe^2}{1-\xe} = \frac{1}{\nH\,\alpha^{\rm rec}_B(\Te)\,\left(A_{2,1}\,\beta^{\rm eff}_{2,1}+A_{2s}\right)} \\
    & \times \left(\int\,\rmd\nu\,\frac{F_\nu(\nu)}{h\nu}\,\sigma^{\rm pi}_2(\nu)\right)\,\left(\int\rmd\nu\,\frac{F_\nu(\nu)}{h\nu}\,\sigma_{\rm Ra}(\nu)\,\frac{f_{\rm Ra}(\nu)}{\tau_{\rm Ra}(\nu)}\right). \nonumber
\end{align}
Here $A_{2,1}=4.695 \times 10^8\,{\rm s}^{-1}$ is the Einstein coefficient for radiative transition from $n=2$ to $n=1$, $A_{2s}=8.229\,{\rm s}^{-1}$ is the two-photon decay rate of the $2s$ state, $\sigma^{\rm pi}_2(\nu)$ is cross section of photo-ionization from $n=2$, $\sigma_{\rm Ra}(\nu)$ is the total cross section of Raman scattering from $n=1$, and $\tau_{\rm Ra}(\nu)$ is the corresponding Raman optical depth of the slab.

\refeq{xe_lowTe} shows that in this low $\Te$ limit $\xe$ is nonzero only if Raman scattering is included. Furthermore, for a fixed shape of $F_\nu(\nu)$ and a fixed ratio between the incident flux and density, $\xe$ is higher for a higher density. An increased slab thickness $d$ lowers the average $\xe$ in the slab because $f_{\rm Ra}(\nu)/\tau_{\rm Ra}(\nu)$ is reduced, which however may be partially compensated due to a decreased Ly$\alpha$ escape probability $\beta^{\rm eff}_{2,1}$.

To quantify the radiation-to-gas ratio, we define a Balmer ``ionization parameter''
\begin{align}
    \mathcal{U}_{\rm BaC} = \frac{\Phi_{\rm BaC}}{\nH\,c} = \frac{1}{\nH\,c}\int_{\rm BaC}\,\rmd\nu\,\frac{F_\nu(\nu)}{h\nu},
\end{align}
where incident photons are integrated between the Balmer limit at $h\nu=3.4\,$eV and the Lyman limit at $13.6\,$eV. \refeq{xe_lowTe} then becomes
\begin{align}
\label{eq:xe_lowTe_UBaC}
     \frac{\xe^2}{1-\xe} = & \frac{\mathcal{U}^2_{\rm BaC}\,\nH\,c^2}{\alpha^{\rm rec}_B(\Te)\,\left(A_{2,1}\,\beta^{\rm eff}_{2,1}+A_{2s}\right)} \\
    & \times\left(\int\,\frac{\rmd\nu}{\Phi_{\rm BaC}}\,\frac{F_\nu(\nu)}{h\nu}\,\sigma^{\rm pi}_2(\nu)\right)\nonumber\\
    & \times \left(\int\frac{\rmd\nu}{\Phi_{\rm BaC}}\,\frac{F_\nu(\nu)}{h\nu}\,\sigma_{\rm Ra}(\nu)\,\frac{f_{\rm Ra}(\nu)}{\tau_{\rm Ra}(\nu)}\right), \nonumber
\end{align}
where the two integrals only depend on the spectral shape of the radiation but not on the absolute flux.

\begin{figure}[ht!]
    \centering
    \includegraphics[scale=0.48]{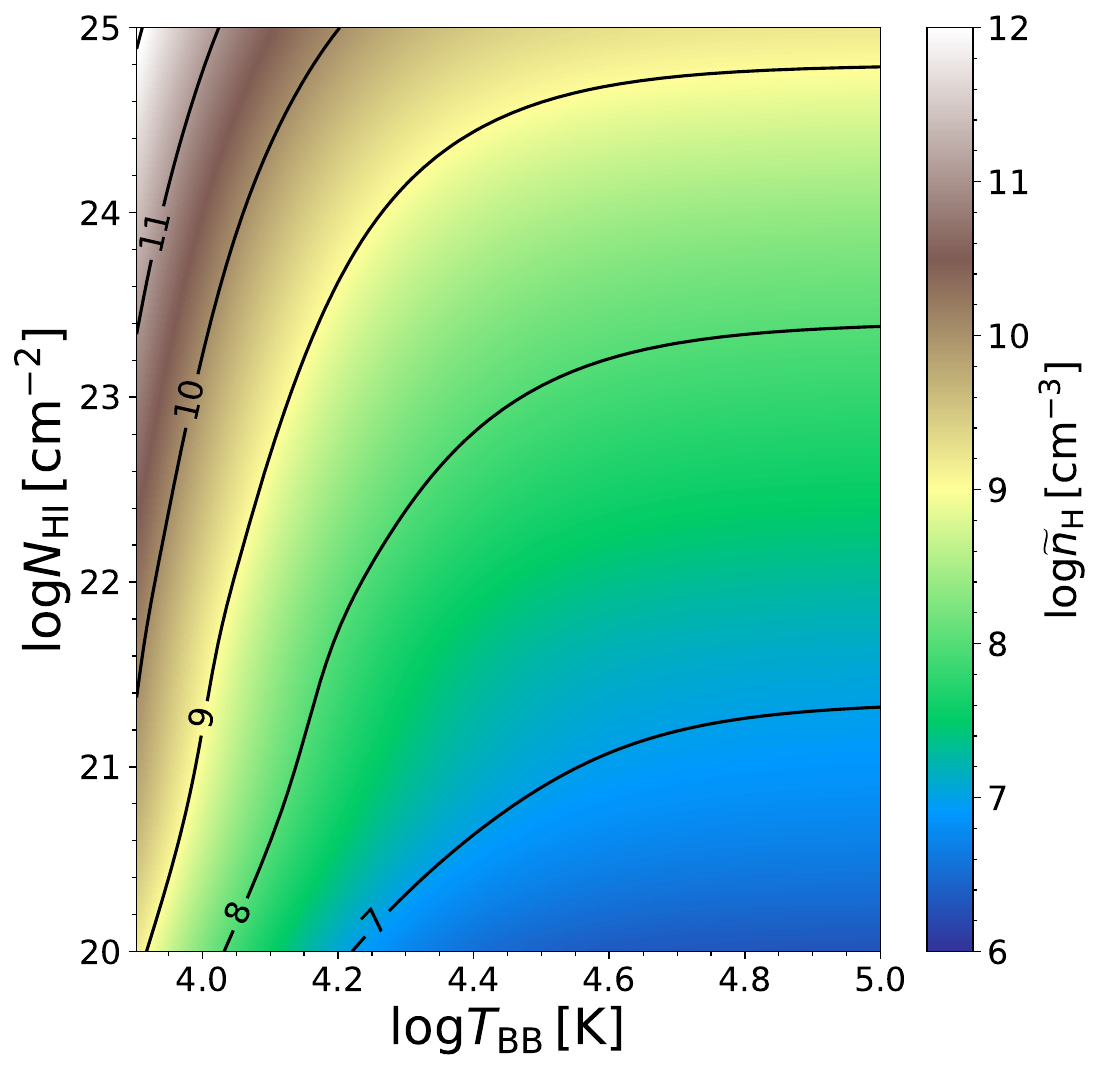}
    \caption{Contours for the reduced density variable $\widetilde n_{\rm H}=\nH\,\mathcal{U}^{-2}_{\rm BaC}\,[90\,\xe^2/(1-\xe)]\,(\alpha^{\rm rec}_{\rm B}(\Te)/4.5\times 10^{-13}\,{\rm cm}^3\,{\rm s}^{-1})$, where the chosen fiducial rate for Case B recombination corresponds to $\Te=5000\,$K. This is computed for the low temperature limit using \refeq{xe_lowTe}, assuming that $2s$ two-photon decay dominates the loss of Ly$\alpha$, $A_{2,1}\,\beta^{\rm eff}_{2,1} \ll A_{2s}$, and as a function of the H I column density $N_{\rm HI}$ and of the temperature $T_{\rm BB}$ of an irradiating blackbody source. Thus, $\widetilde n_{\rm H}$ is the density when $\mathcal{U}_{\rm BaC}=1$ and $\xe=0.1$. Since the dependence on $\mathcal{U}_{\rm BaC}$ and $\xe$ are explicit from \refeq{xe_lowTe_UBaC}, $\nH$ for any other values of $\mathcal{U}_{\rm BaC}$ and $\xe$ can be deduced from this plot.}
    \label{fig:nH_contour_plot}
\end{figure}

If the Ly$\alpha$ optical depth is large $\tau_{{\rm Ly}\alpha} \gtrsim 3 \times 10^7$, corresponding to $N_{\rm HI} \gtrsim 3.6\times 10^{20}\,{\rm cm}^{-2}\,(\Te/5000\,{\rm K})^{1/2}$, the slab loses Ly$\alpha$ photons primarily through $2s$ two-photon decay, so that $A_{2,1}\,\beta^{\rm eff}_{2,1} + A_{2s} \approx A_{2s}$. \reffig{nH_contour_plot} allows us to determine the relation between $\nH$ and $\xe$, for a blackbody incident spectrum at a spectral temperature $T_{\rm BB}$ and as a function of $N_{\rm HI}=(1-\xe)\,\nH\,d$. For a fixed spectral shape and a fixed $\mathcal{U}_{\rm BaC}$, a higher ionization $\xe$ requires a higher density $\nH$, which explains the trend we will see later in \reffig{slab_UBaC}.

\refeq{xe_lowTe_UBaC} shows that a larger $\mathcal{U}_{\rm BaC}$ increases ionization. In reality, $\mathcal{U}_{\rm BaC}\gg 1$ would be limited by the effect of radiation pressure, just like the dynamic limit $\mathcal{U} \lesssim 0.1$ for the LyC ionization parameter of the H II zone~\citep{YehMatzner2012RadiationWindFeedback}. Radiation pressure can come either from ionizing photons absorbed in the H II zone in front of the slab, from Rayleigh and Raman scatterings of the incident FUV photons, or from dust absorption. Once the radiation pressure is too strong, $\nH$ increases as the gas slab is compressed.

\subsection{Numerical Results for Weigelt Blobs}
\label{sec:numres}

Since $\eta$ Car and its nebula have been well observed over decades at good spatial resolution, we consider a situation similar to the Weigelt blobs. There are strong evidences that the $\eta$ Car system is a binary star with a 5.5-yr period~\citep{Hillier2001EtaCar}. The primary star is a post-eruption LBV with $L_{{\rm bol},{\rm A}}=5\times 10^6\,L_\odot$, hiding behind an optically thick stellar wind (mass loss rate $\dot M_{w,{\rm A}} \lesssim 10^{-3}\,M_\odot\,{\rm yr}^{-1}$~\citep{Clementel2015EtaCar3DWindCollisionModel}) with a temperature $T_{\rm A} \approx 10000\,$K at the continuum photosphere~\citep{Hillier2001EtaCar}. We will therefore use a blackbody spectrum for its photospheric emission. The secondary star is an O star, with $L_{{\rm bol},{\rm B}}\simeq 4 \times 10^5\,L_\odot$ and $T_{\rm B} \approx 39000\,$K~\citep{Mehner2010a}. The secondary star provides essentially all the ionizing photons, $\log Q({\rm H^0})[{\rm s}^{-1}]=49.5$, and dominates the FUV radiation blueward of $1500\,\AA$. The primary star dominates the NUV band $3000$--$4000\,\AA$. The corresponding Lyman and Balmer ionization parameters are
\begin{align}
    \mathcal{U} & = 0.015\,\left(\frac{\nH}{10^8\,{\rm cm}^{-2}}\right)^{-1}\,\left(\frac{D}{500\,{\rm AU}}\right)^{-2}, \\
    \mathcal{U}_{\rm BaC} & = 0.58\,\left(\frac{\nH}{10^8\,{\rm cm}^{-2}}\right)^{-1}\,\left(\frac{D}{500\,{\rm AU}}\right)^{-2}.
\end{align}

We set a gas slab at a fiducial distance $D =500\,{\rm AU}$ from the binary as inferred astrometrically from data~\citep{Hillier2001EtaCar}. The slab has a thickness
\begin{align}
    d=21\,{\rm AU}\,\left( \frac{N_{\rm H}}{10^{22.5}\,{\rm cm}^{-2}}\right)\,\left(\frac{\nH}{10^8\,{\rm cm}^{-3}}\right)^{-1}.
\end{align}
LyC radiation pressure acting on the H II zone is 
\begin{align}
\label{eq:PLyC}
    P_{\rm LyC}=3.5\times 10^{11}\,{\rm K}\,{\rm cm}^{-3}\,\left( \frac{D}{500\,{\rm AU}}\right)^{-2}.
\end{align}

We parameterize the dust absorption cross section per hydrogen by normalizing it at the Ly$\alpha$ wavelength $\sigma_{\rm d}(\lambda_\Lya) = 1.4 \times 10^{-21}\,f_{\rm d}\,{\rm cm}^2$, where $f_{\rm d}=1$ corresponds to a dust-to-gas ratio typical for solar metallicity. For the wavelength dependence of grain absorption, we use the Milky Way grain model with $R_V=5.5$ in \cite{Weingartner2001DustGrainModelMWLMCSMC}.

\begin{figure*}[ht!]
    \centering
    \includegraphics[scale=0.57]{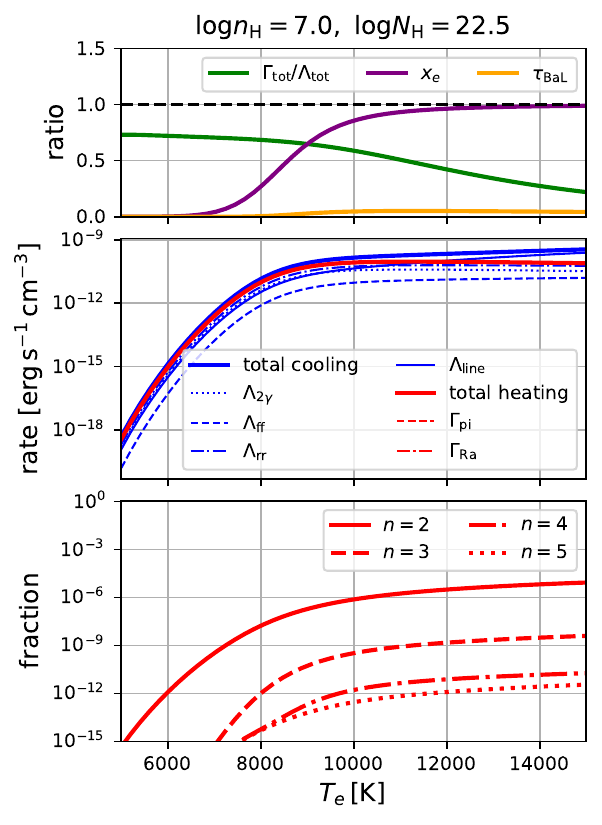}
    \includegraphics[scale=0.57]{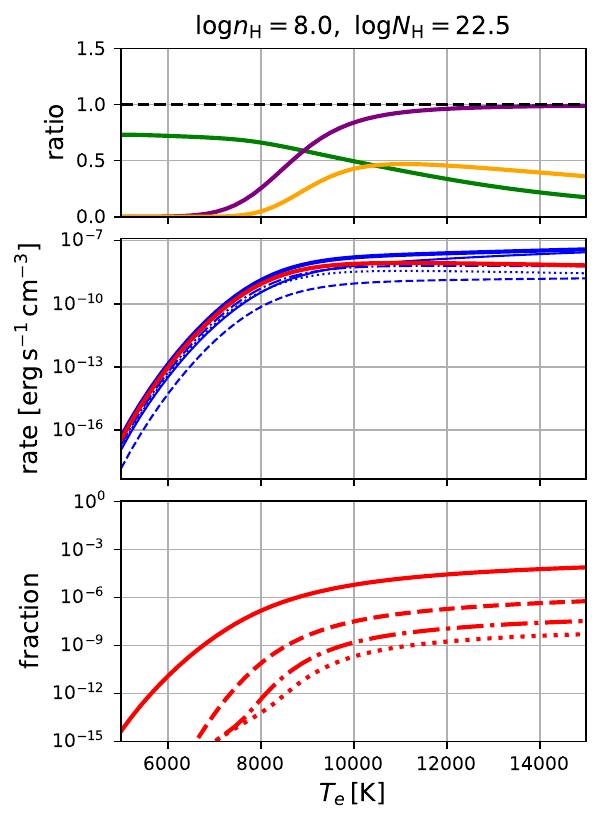}
    \includegraphics[scale=0.57]{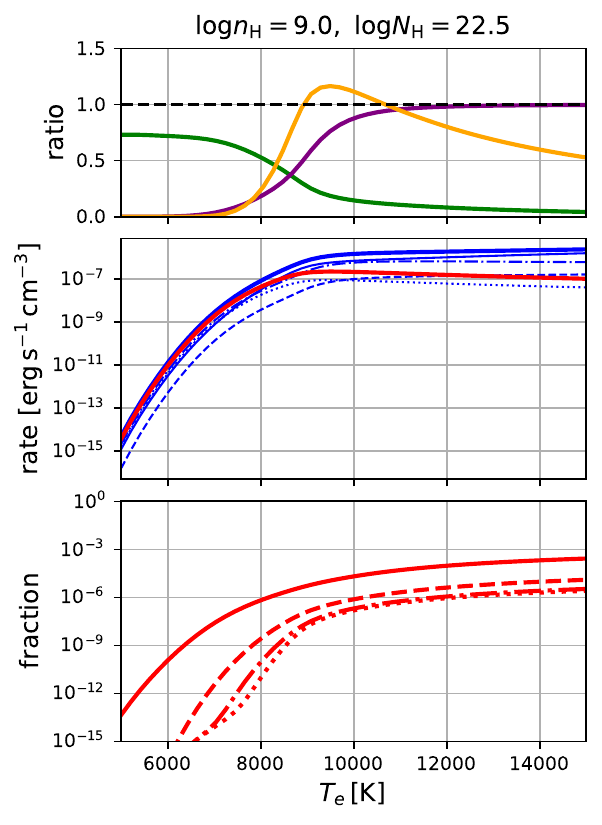}
    \caption{Results for the one-zone, dust-free hydrogen gas slab, and without accounting for Raman scattering of incident continuum photons. The calculation is done for irradiation by the $\eta$ Car binary system from a distance $D=500\,$AU. The columns correspond to three density values $\nH$ and the same column density $N_{\rm H}$. The top panels show the ratio between energy gain and loss $\Gamma_{\rm tot}/\Lambda_{\rm tot}$ (green), ionization fraction $\xe$ (magenta), and the optical depth at the Balmer continuum limit $\tau_{\rm BaL}$ (orange). The middle panels show volumetric rates of energy gain (red): photo-ionization $\Gamma_{\rm pi}$, Raman scattering $\Gamma_{\rm Ra}$; and rates of energy loss (blue): recombination $\Lambda_{\rm rr}$, free-free emission $\Lambda_{\rm ff}$, H $2s$ two-photon decay $\Lambda_{2\gamma}$, and escaping resonant lines $\Lambda_{\rm line}$. The bottom panels show the fractional populations for the H I excited states. In these examples, the total heating rate falls short of the total cooling rate in the temperature range $\Te=5000$--$15000\,$K and no equilibrium is possible.}
    \label{fig:slab_noRa}
\end{figure*}

For given values of $\nH$ and $N_{\rm H}=\nH\,d$, we calculate the steady-state ionization fraction $\xe$ and level populations as a function of $\Te$. We explore a range of gas density $\log\nH=7$--$9$ appropriate for the Weigelt blobs.  

As shown in \reffig{slab_noRa}, when energy is injected into the slab only through photo-ionization, heating does not keep up with cooling across a wide temperature range $\Te=5000$--$15000\,$K we calculate, and hence the absence of equilibrium. While the cooling rate exponentially drops as $\Te$ decreases below $8500\,$K, the heating rate drops rapidly too and is always lower. This reflects the inability of the slab to sustain a significant $n=2$ (and higher $n$) population, which is required to absorb enough radiation energy. 

\begin{figure*}[ht!]
    \centering
    \includegraphics[scale=0.57]{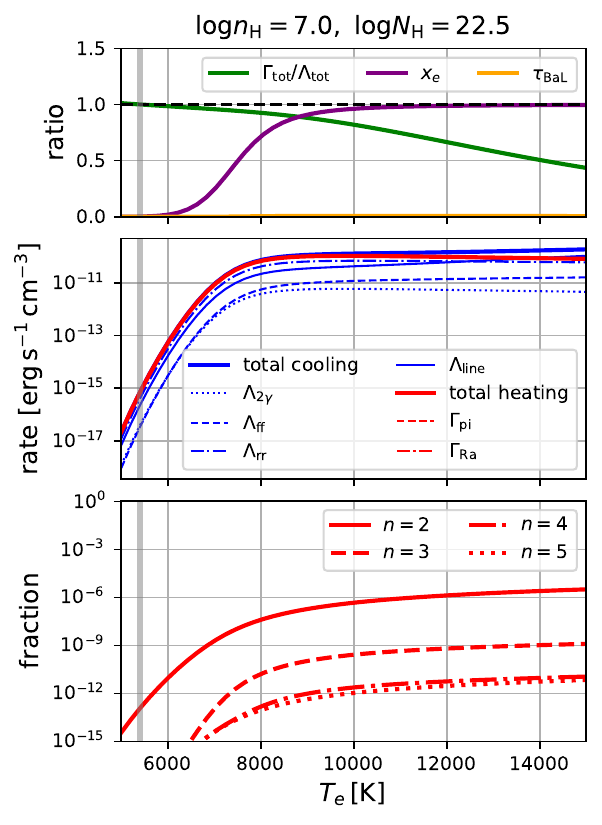}
    \includegraphics[scale=0.57]{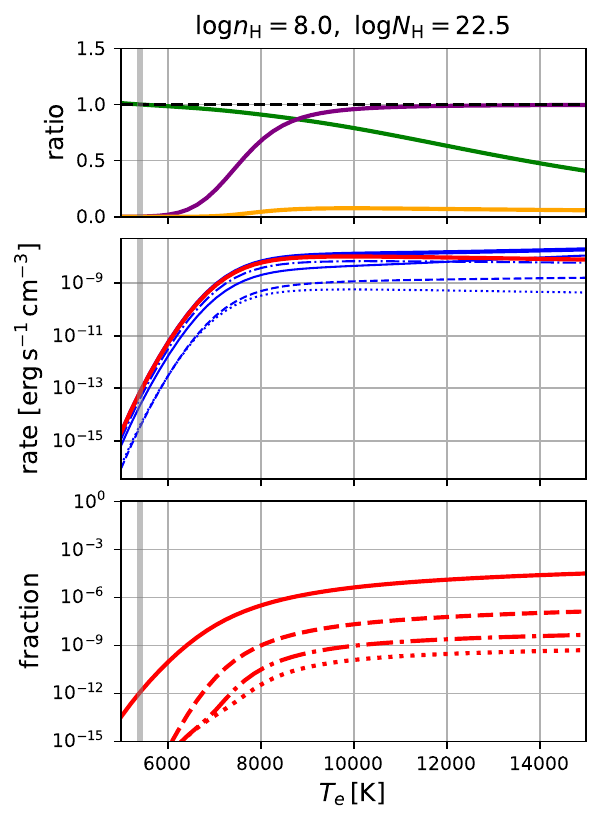}
    \includegraphics[scale=0.57]{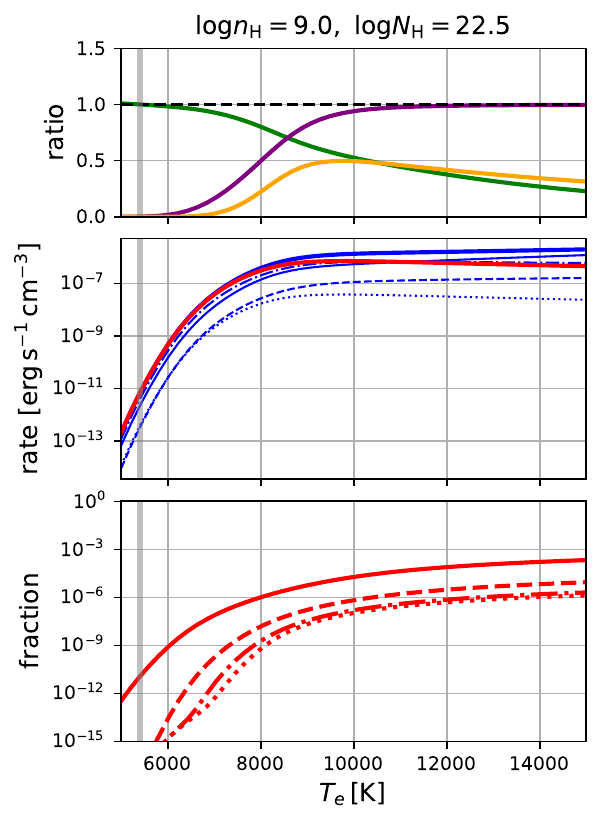}
    \caption{Same as \reffig{slab_noRa} but with incident continuum flux increased by eight fold. This corresponds to decreasing the distance of the gas to the continuum source $D$ by about a factor of three. A stable equilibrium in which heating balances cooling arises at $\Te\approx 5400\,$K in all three panels, albeit with rather small $\xe$ values.}
    \label{fig:slab_noRa_rad8x}
\end{figure*}

If incident radiation increases by 8 fold, an equilibrium arises at $\Te\approx 5400\,$K (\reffig{slab_noRa_rad8x}). This would correspond to a reduced distance $D=177\,$AU for the Weigelt blobs, which however seems too small to be reality.

The limitation here has to do with the need to maintain a steady state of internal Ly$\alpha$ radiation. To keep the gas partially ionized through photo-ionization, Ly$\alpha$ photons are required for pumping a significant $n=2$ population. For an open geometry, Ly$\alpha$ continuously escape the slab, or get destroyed via other pathways, which has to be compensated for by replenishment mechanisms. Note that photo-ionization from the excited states of hydrogen is not a net source of Ly$\alpha$ photons, unlike that from the ground state. Electron collisional excitation of $n=2$ from the ground state is a source of Ly$\alpha$ photons, but is only efficient if the equilibrium $\Te$ is high enough, at which cooling overtakes heating. This highlights the importance of the fluorescent source of Ly$\alpha$ photons, through Raman-scattering on the damping wings of the Lyman lines. To capture incident continuum photons, Raman scattering only requires a large hydrogen column in the ground state, but does not require a high $\Te$.

\begin{figure*}[ht!]
    \centering
    \includegraphics[scale=0.57]{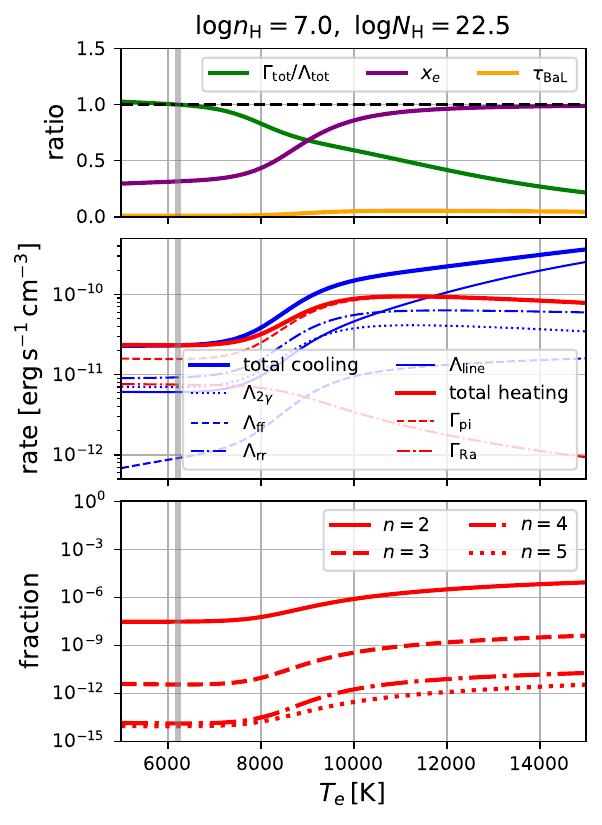}
    \includegraphics[scale=0.57]{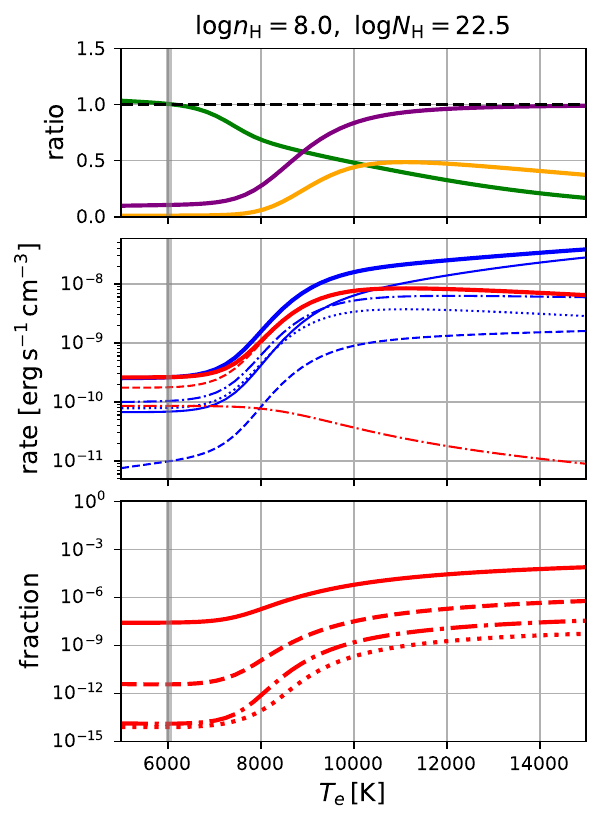}
    \includegraphics[scale=0.57]{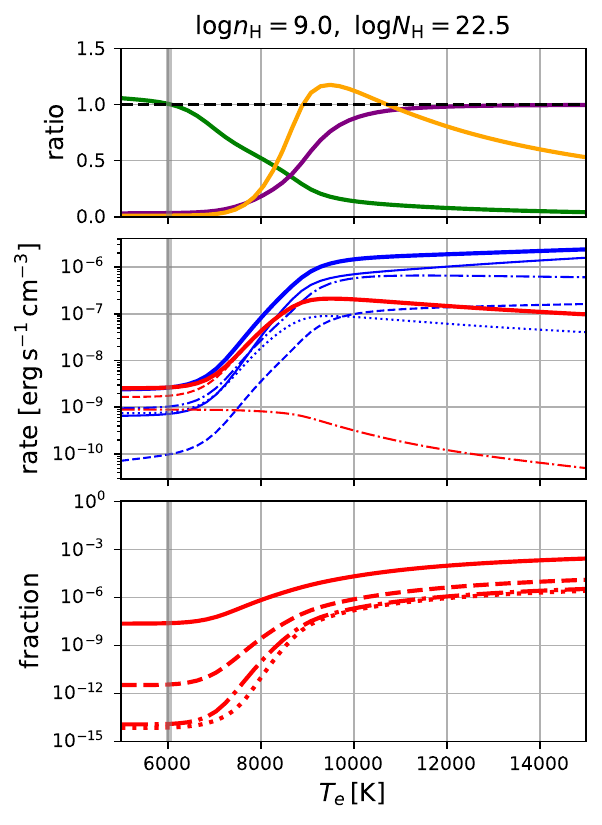}
    \caption{Same as \reffig{slab_noRa} but accounting for Raman scattering of incident continuum photons. A stable equilibrium is found (vertical grey line) at $\Te\approx 6000~$K when heating balances cooling $\Gamma_{\rm tot}/\Lambda_{\rm tot} = 1$. At this equilibrium, the gas has a significant ionization fraction and a significant $n=2$ fractional population $x_2 \simeq 10^{-8}$--$10^{-7}$.}
    \label{fig:slab}
\end{figure*}

In \reffig{slab}, we perform calculations accounting for Raman scattering using the framework introduced in \refsec{Raman}. In the circumstellar environment of $\eta$ Car, UV continuum photons are copiously provided. For $\log\nH=7$--$9$, a stable equilibrium is found at a warm temperature $\Te\approx 6000\,$K. The equilibrium state is also partially ionized, and has a significant fraction of $n=2$ atoms, $x_2\simeq 10^{-8}$--$10^{-7}$. The comparison between \reffig{slab} and \reffig{slab_noRa} demonstrates that Raman scattering crucially enables a partially ionized, warm equilibrium state without the need for external LyC photons. Consistent with our assumption, the slab at equilibrium has a small optical depth at the Balmer limit $\tau_{\rm BaL} \ll 1$ and blueward of it.

Raman scattering also causes the slab to gain energy. This ultimately heats the gas as a super-thermal $n=2$ population is established for which electron collisional (de-)excitation has a net effect of heating the gas. However, at equilibrium photo-ionization still dominates heating. What happens is that Raman scattering leverages the $n=2$ population through the resultant Ly$\alpha$ supply, which allows the slab to absorb more incident UV photons. 

In a more realistic situation, we expect that Ly$\alpha$ photons are produced copiously in the H II zone in the front, and some of those can diffuse into the partially ionized zone. This is in principle another source of Ly$\alpha$, but is likely inefficient in pumping the $n=2$ population, as the large scattering optical depth into the partially ionized zone renders diffusivity rather small.

\begin{figure*}[ht!]
    \centering
    \includegraphics[scale=0.57]{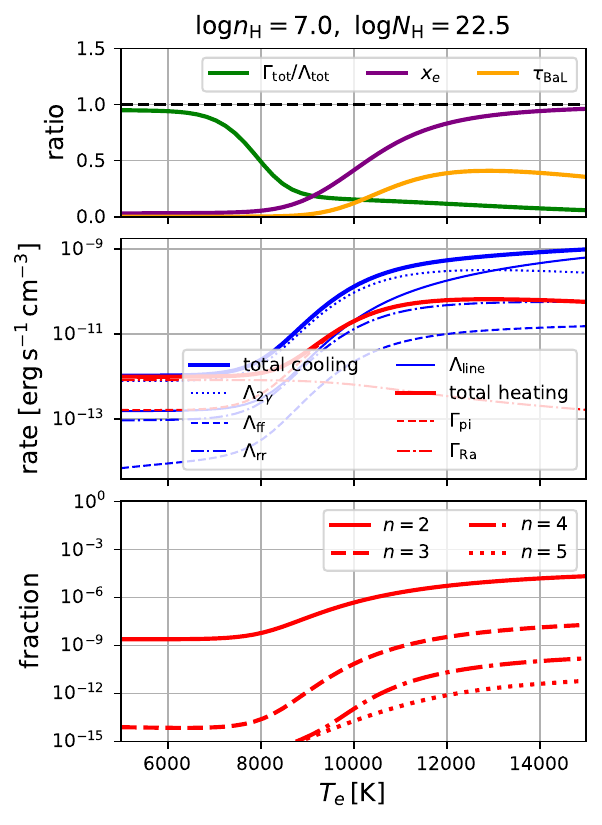}
    \includegraphics[scale=0.57]{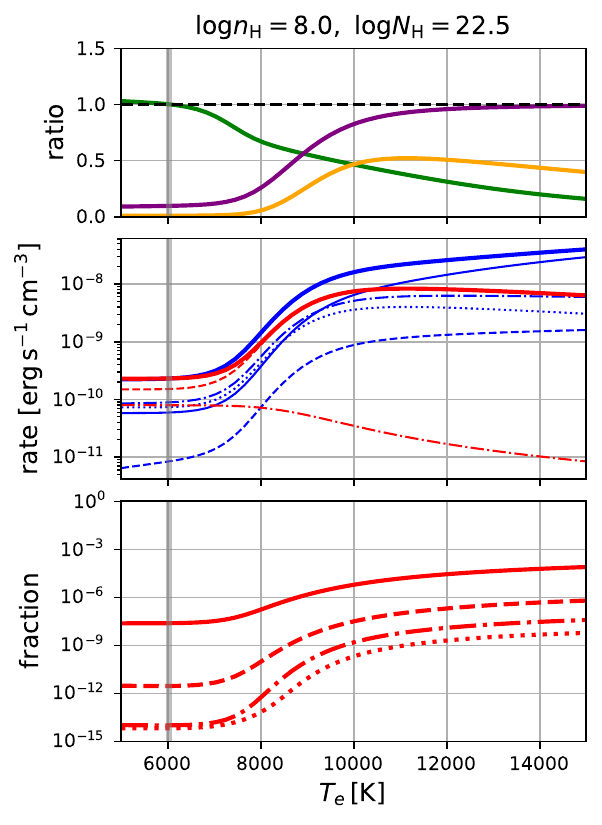}
    \includegraphics[scale=0.57]{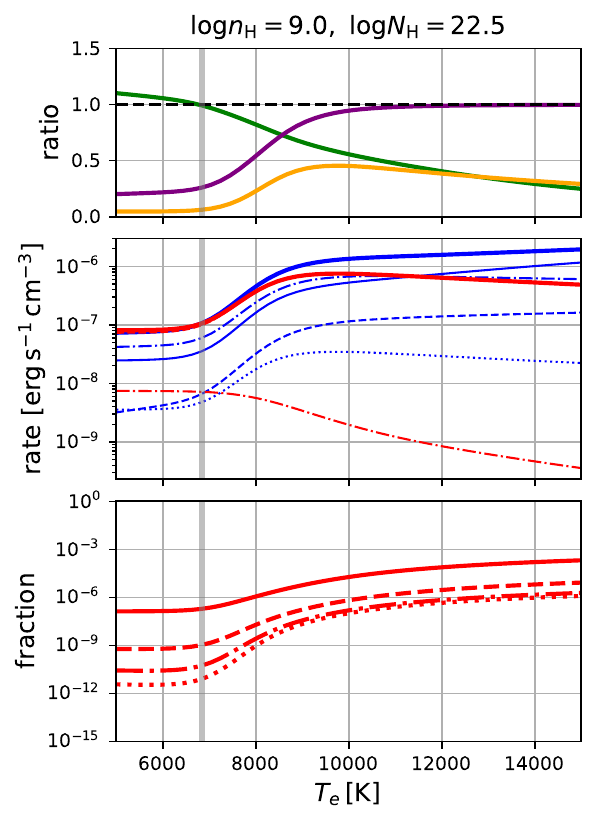}
    \caption{Same as \reffig{slab} but the incident Balmer continuum photon flux is set to be proportional to the gas density, $\mathcal{U}_{\rm BaC} = \Phi_{\rm BaC}/(\nH\,c)=0.5$. Equilibrium is more easily realized at higher $\nH$, with a larger ionization fraction $\xe$ and a warmer equilibrium temperature $\Te$. The $\log \nH=7$ model does not allow an equilibrium with $\Te>5000\,$K.}
    \label{fig:slab_UBaC}
\end{figure*}

In \reffig{slab}, the equilibrium $\xe$ decreases with increased $\nH$. This is because we have fixed the incident stellar radiation flux. \reffig{slab_UBaC} demonstrates a test where we scale the incident radiation in proportion to $\nH$, by setting $\mathcal{U}_{\rm BaC}=0.5$. This roughly corresponding to $\log\nH=8$ for the Weigelt blobs for $D=500\,$AU. Equilibrium does not exist for $\log\nH=7$, but is found for $\log\nH=8$ and $\log\nH=9$. At fixed $\mathcal{U}_{\rm BaC}$, the equilibrium has a higher $\xe$ and a warmer $\Te$ as $\nH$ increases, aligned with our discussion of the low temperature limit in \refsec{lowTe}. These results imply that significant partial ionization is impossible at much lower densities $\log \nH \lesssim 4$ found in ordinary H II regions powered by young stars.

\begin{figure*}[ht!]
    \centering
    \includegraphics[scale=0.53]{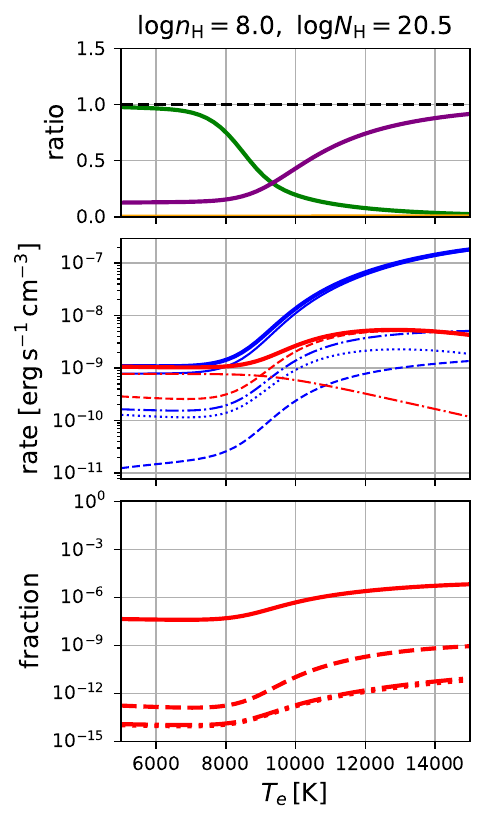}
    \includegraphics[scale=0.53]{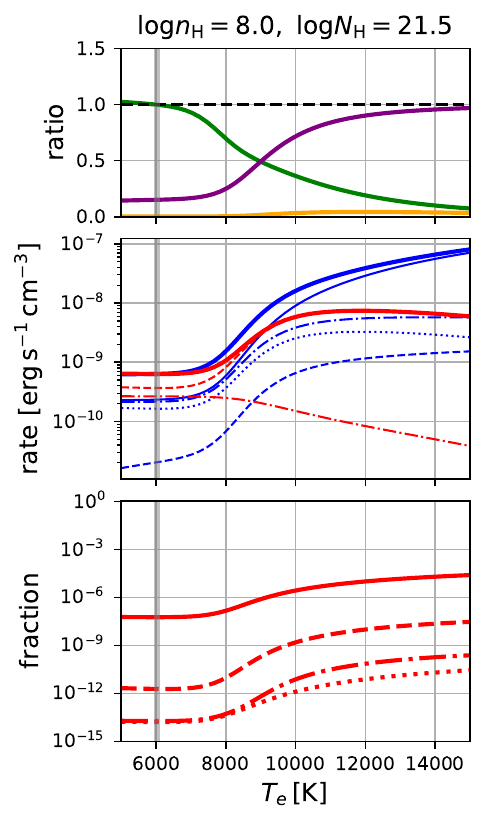}
    \includegraphics[scale=0.53]{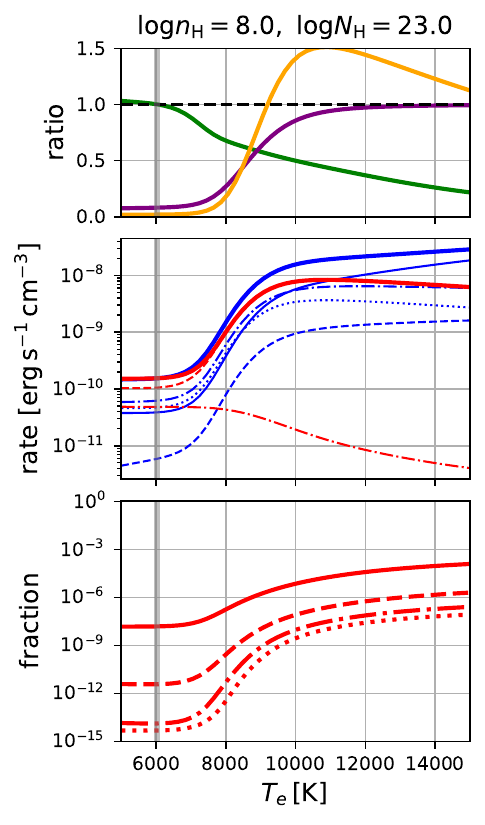}
    \includegraphics[scale=0.53]{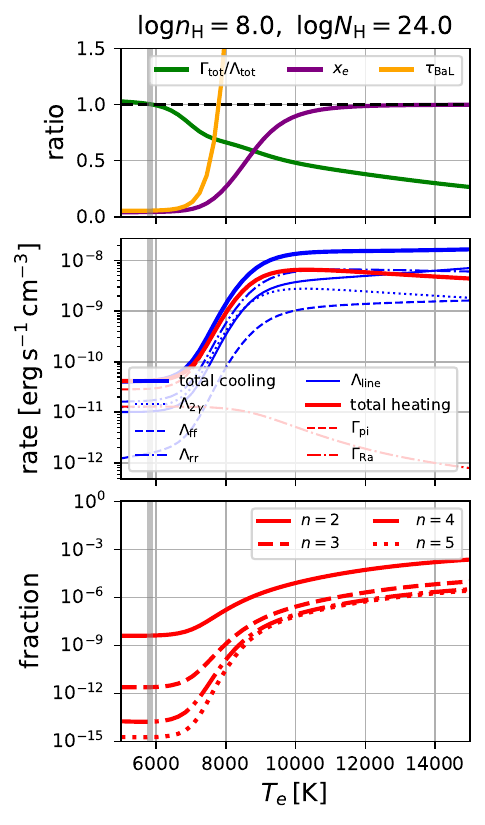}
    \caption{Same as \reffig{slab} but for a fixed $\log \nH=8$ but 4 different hydrogen column densities $N_{\rm H}$, in addition to the $\log N_{\rm H}=22.5$ case which is already shown in the middle column of \reffig{slab}. For the very thin slab with $\log N_{\rm H}=20.5$, cooling due to escaping line photons is too efficient, and no equilibrium is found down to $\Te=5000\,$K. When the slab is too thick, such as with $\log N_{\rm H}=24$, growth of Ly$\alpha$ injection does not keep up with the increase of hydrogen mass, as the slab is starved of incident continuum photons that can be Raman absorbed, so at equilibrium the ionization fraction is rather low. For a realistic irradiated slab with a stratified structure, partial ionization will weaken with increased depth, as can be seen in the \texttt{CLOUDY} models in \reffig{cloudy_models}.}
    \label{fig:slab_NHtest}
\end{figure*}

In \reffig{slab_NHtest}, we study the dependence on the column density $N_{\rm H}$. For a thin column, Raman scattering is limited to continuum photons in the very vicinity of Ly$\beta$, Ly$\gamma$, and so on, but the total hydrogen mass to partially ionize is also small. The other important effect is that resonant line trapping is inefficient at low $N_{\rm H}$, so line cooling is increased. The model with $\log N_{\rm H}=20.5$ shown in \reffig{slab_NHtest} does not have an equilibrium at $\Te > 5000\,$K. As $N_{\rm H}$ increases, heating can balance cooling and sustain partial ionization. When $N_{\rm H}$ grows too large, absorption of continuum photons by Raman scattering saturates at a given wavelength, so the growth of Ly$\alpha$ injection does not keep up with the increase in the gas mass. As a result, while the $\log N_{\rm H}=24$ model has an equilibrium at $\Te\approx 6000\,$K, the ionization fraction is only a few percent. In realistic situations, gas properties vary with the depth. The pumping of the $n=2$ population decreases with depth, gradually shutting down partial ionization.

\section{CLOUDY Model}
\label{sec:cloudy}

The one-zone calculated in \refsec{one_zone} does not account for helium and metals. In predominantly H I gas, metals contribute to line cooling, while at the same time heats the gas through ionization by UV photons. The one-zone model ignores spatial variation, and simplistically assumes that the mere role the H II region in the front plays is to shield all LyC photons. To improve beyond these limitations, we present photo-ionization calculations using the \texttt{CLOUDY} code~\citep{Ferland2017c17}, taking into account the H II zone and what is behind it, and under irradiation conditions appropriate for the Weigelt blobs (as stated in \refsec{one_zone}).

The models are set to be isobaric with a total pressure $P$. A wind-wind collision structure is predicted for the $\eta$ Car binary system. Since the secondary star is on a tight, highly eccentric orbit ($e \simeq 0.9$ and semi-major axis $a=15.45\,$AU) around the primary, the secondary spends most of the time on the side of the apoastron. The Weigelt blobs probably lie on that side relative to the central binary such that ionizing radiation from the secondary is not obscured by the opaque wind of the primary. For fiducial wind parameters inferred for the secondary~\citep{PittardCorcoran2002EtaCarWindParameters, Parkin2009EtaCarCollidingWinds3DModel}, the ram pressure on the Weigelt blobs is
\begin{align}
P_{\rm ram} & = 2 \times 10^{12}\,{\rm K}\,{\rm cm}^{-3}\,\left(\frac{\dot M_{w,{\rm B}}}{10^{-5}\,M_\odot\,{\rm yr}^{-1}}\right)\nonumber\\
& \times \left(\frac{v_{w,{\rm B}}}{3000\,{\rm km}\,{\rm s}^{-1}}\right)\,\left(\frac{D}{500\,{\rm AU}}\right)^{-2},
\end{align} 
which significantly exceeds the LyC radiation pressure \refeq{PLyC}. This suggests that the Weigelb blobs are hydrodynamically compressed to $\log P \simeq 12$--$12.5$.

We set element abundances according to Table 4 of \cite{Verner2002FePhotoionModel}. Solar abundances are used for elements missing from that table. Weigelt blobs have an unusual chemical composition, as the gas was heavily CNO-processed before ejected by the primary star. The gas is rich in N but deficient in C and O~\citep{Hamann1999HSTSTISspectraEtaCarEjecta, Verner2002FePhotoionModel}. Moreover, \cite{Verner2002FePhotoionModel} suggests a factor of five enrichment of He, which is plausible if the gas originates from an evolved star.

\begin{figure*}[ht!]
    \centering
    \includegraphics[scale=0.45]{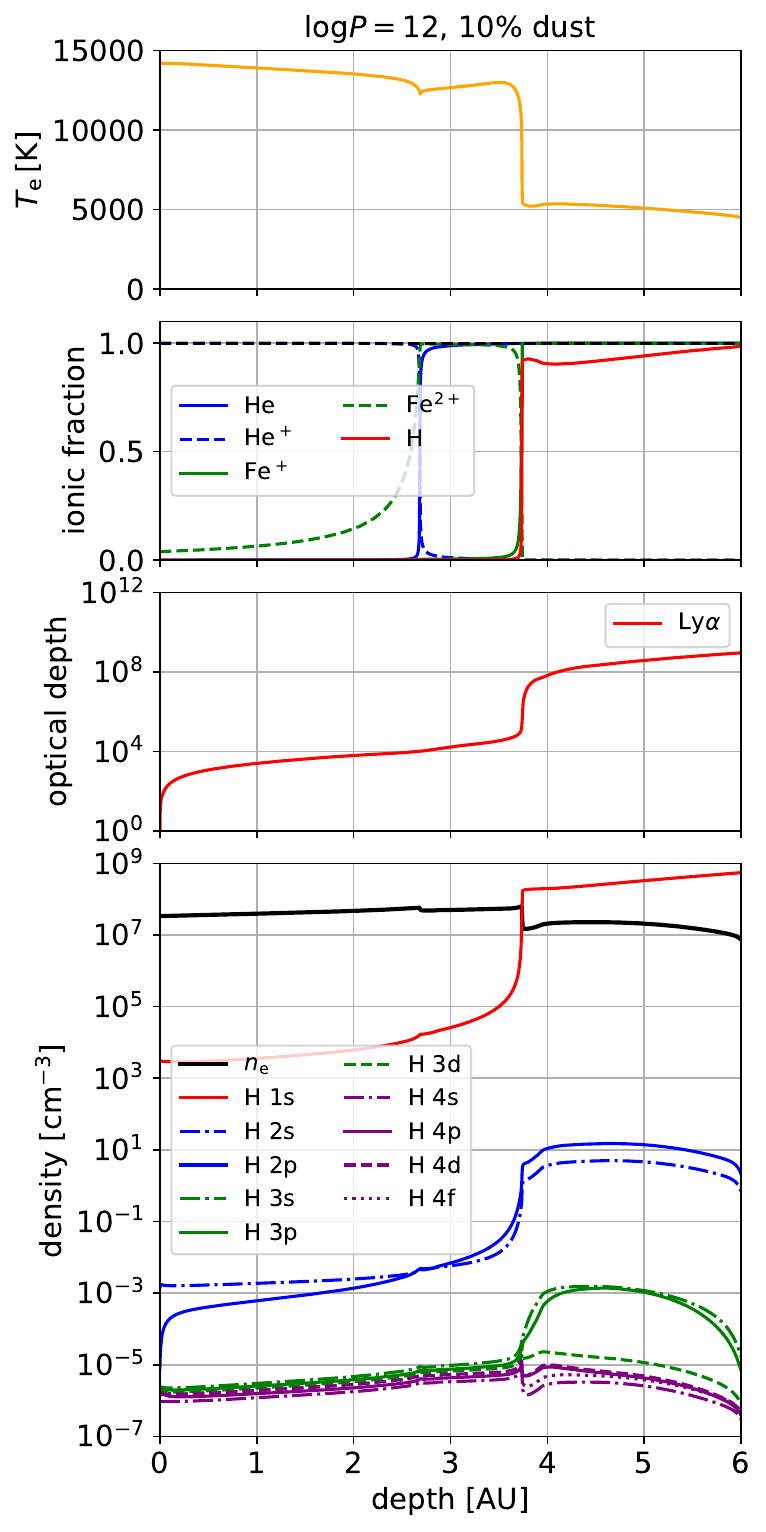}
    \includegraphics[scale=0.45]{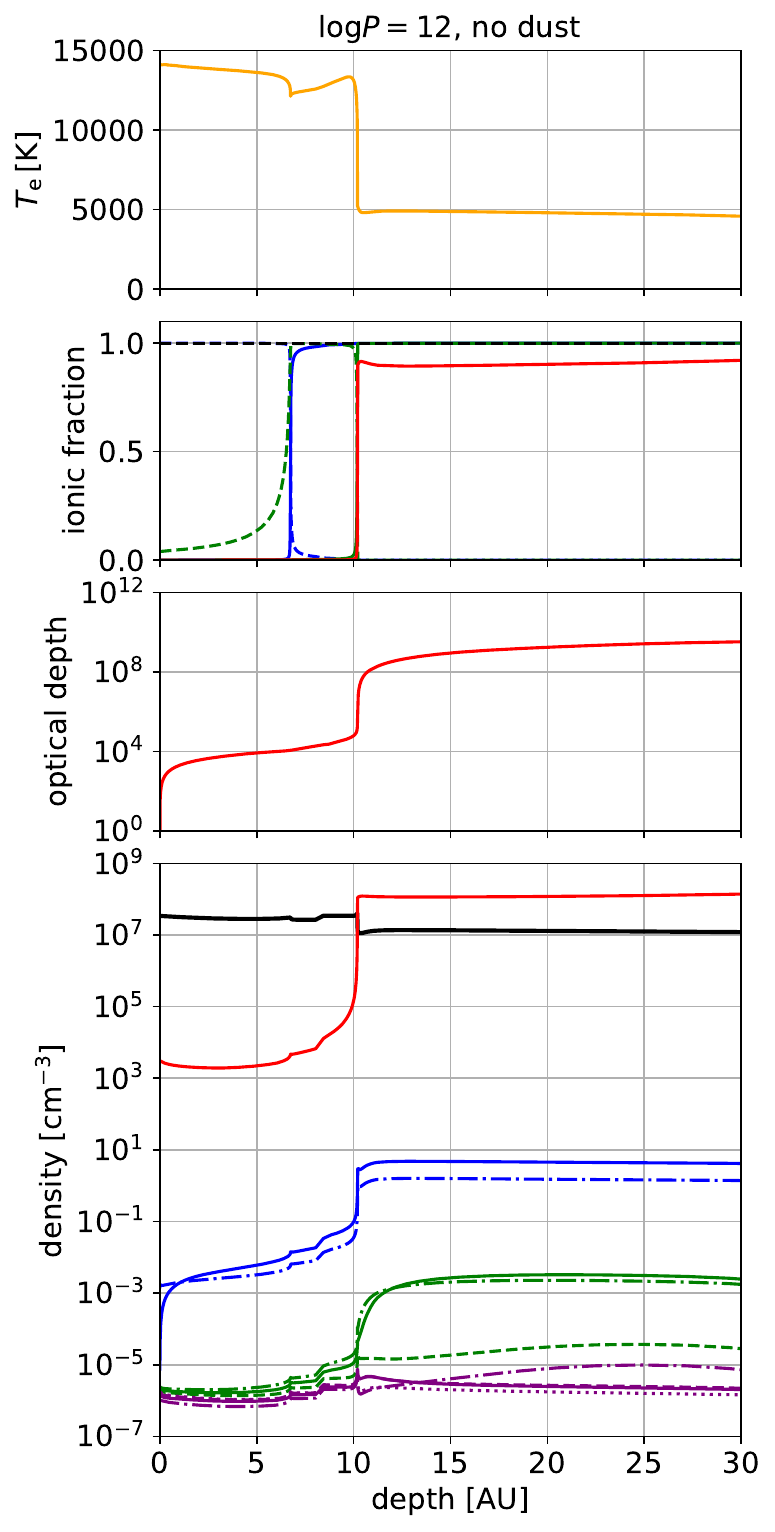}
    \includegraphics[scale=0.45]{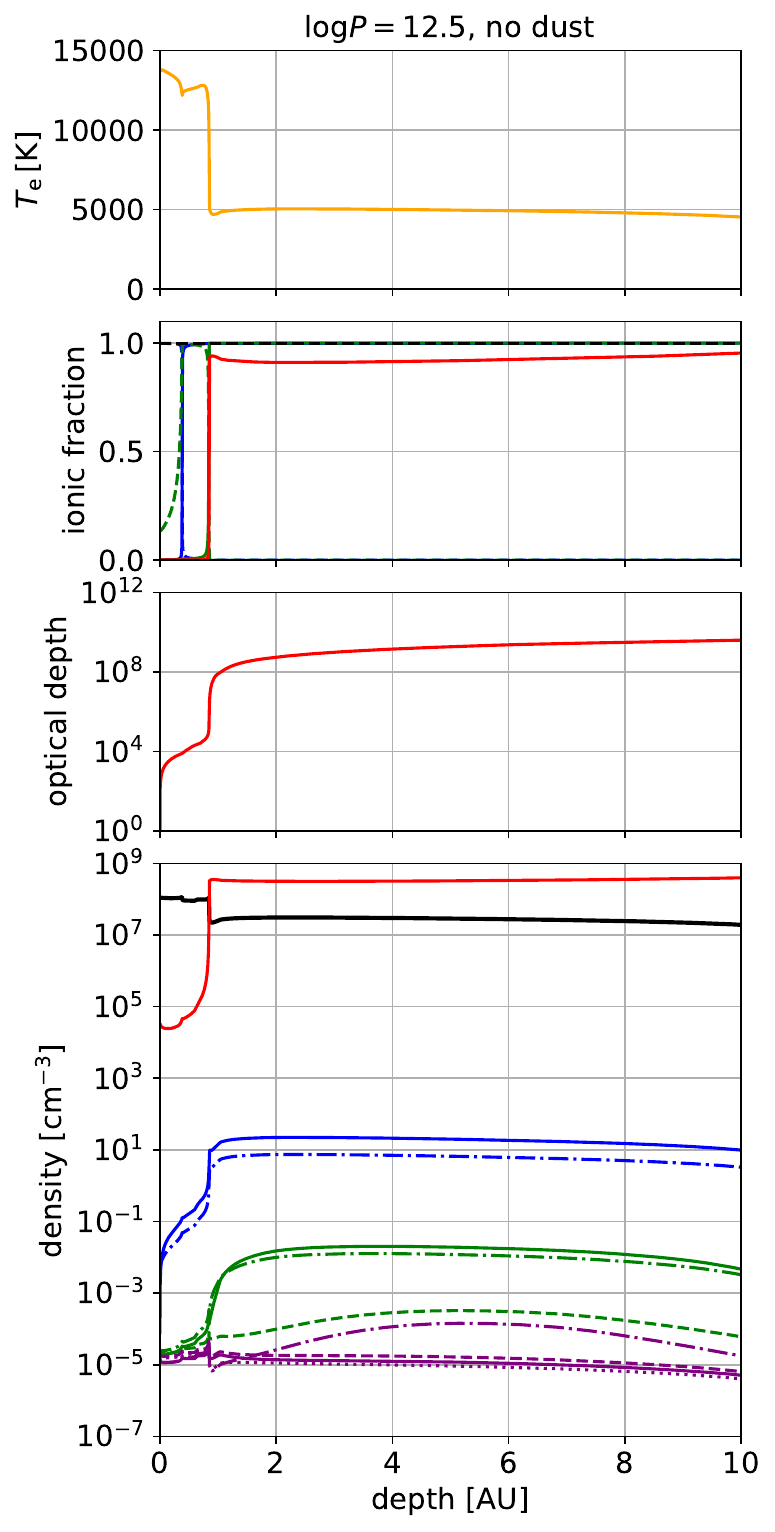}
    \caption{Isobaric photo-ionization models calculated with \texttt{CLOUDY} assuming the radiation condition of the Weigelt blobs of $\eta$ Car and element abundances recommended by \cite{Verner2002FePhotoionModel} (see main text for more details). Gas temperature, ionization, Ly$\alpha$ optical depth, electron density and hydrogen level populations are plotted as a function of depth from the irradiated surface of the gas slab. We show three models (total pressure $P$ in units of ${\rm K}\,{\rm cm}^{-3}$): $\log P=12$ with a dust-to-gas ratio 10\% of that in the typical interstellar gas at solar metallicity (left column), $\log P=12$ without dust (middle column), $\log P=12.5$ without dust (right column). In these models, calculation is performed up to $\log N_{\rm H I}=22$--$22.5$. The gas behind the usual H ionization front has a low level of ionization $\xe\lesssim 10\%$, a lukewarm temperature $\Te \approx 5000\,$K, and an enhanced H I $n=2$ fractional population $x_2 \sim 10^{-7}$. Due to large Balmer line optical depths, the $n=3$ atoms are $\sim 2$ orders of magnitude more abundant than the $n=4$ atoms, unlike in the H II zone. This causes a large H$\alpha$/H$\beta$ flux ratio. Due to the five-fold elevated He abundance assumed, an intermediate zone of ionized H and neutral He forms, where ionic species of intermediate ionization potential such as Fe III are abundant.}
    \label{fig:cloudy_models}
\end{figure*}

In \reffig{cloudy_models}, we show the 1D structure of the nebula, as a function of depth from the irradiated surface at $D=500\,$AU from $\eta$ Car. Just below the surface, there is the zone with ionized H and singly ionized He. A zone of ionized H but neutral He lies behind it, which results from the 5-fold enhanced He abundance. This intermediate zone, with accounts for an order-unity fraction of the H II zone, hosts ionic species of intermediate ionization potential, such as Fe III, which may have implications for the observed strong fluorescent Fe III lines~\citep{Johansson2000FeIIIforbiddenlines, Zethson2012Weigeltlines}. This intermediate zone is bounded by the H ionization front at which incident LyC photons are depleted. 

Beyond this front, the gas exhibits a low ionization fraction $\xe \lesssim 10\%$ and is kept at $\Te \approx 5000\,$K. A trend is seen that partial ionization weakens as the depth increases. The degree of ionization is somewhat lower than found in the comparable one-zone, hydrogen slab model (\refsec{one_zone}). We note that \texttt{CLOUDY} might have underestimated excitation to $n=2$ (and higher excited states) through the Raman processes, as the source integral is artificially cutoff at about ten times the line Doppler width; beyond that, only Rayleigh scattering is included in the continuum opacity (private comm. with G. Ferland).

\begin{figure*}[ht!]
    \centering
    \includegraphics[width=\textwidth]{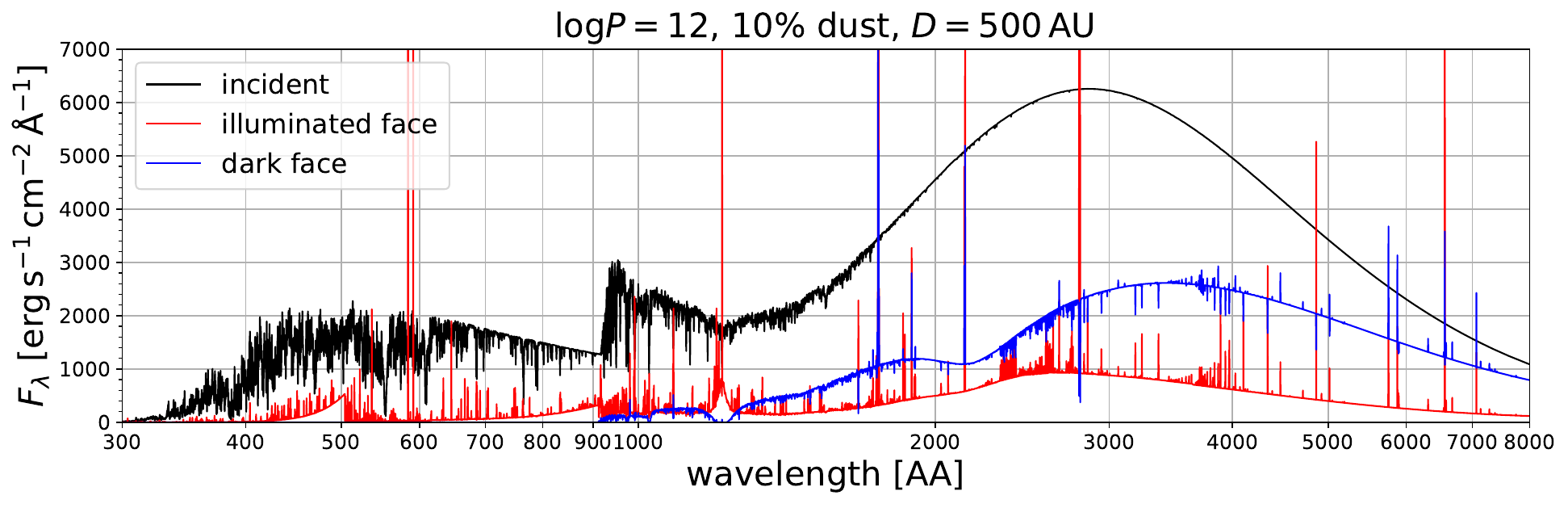}
    \includegraphics[width=\textwidth]{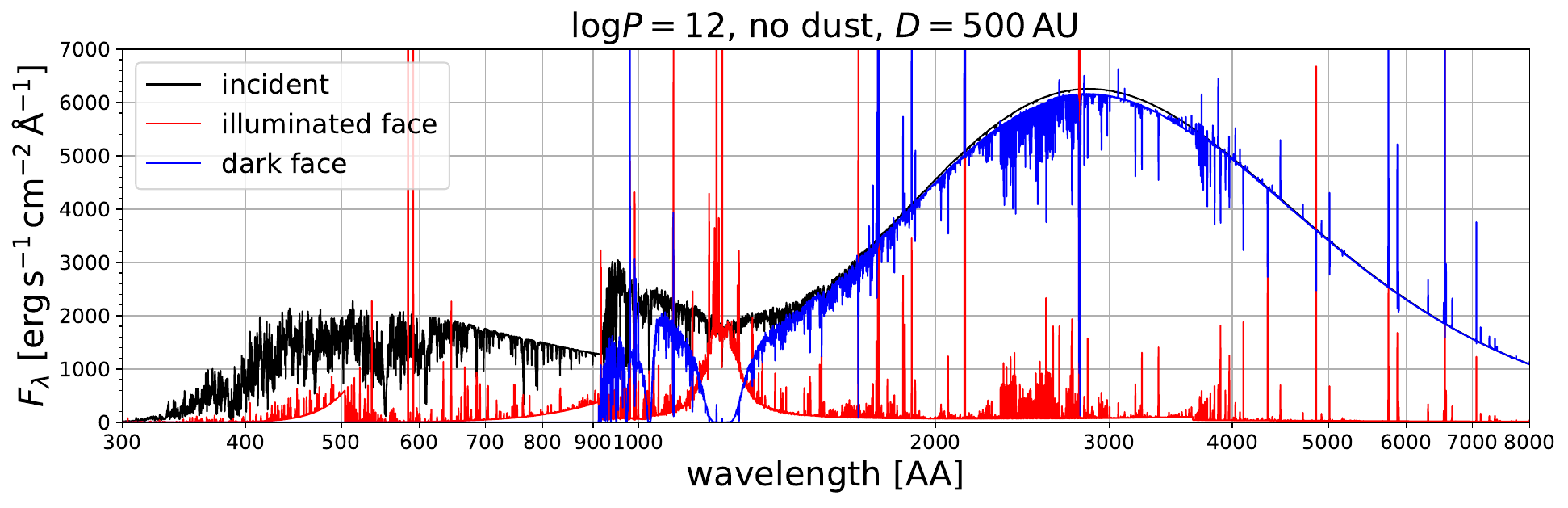}
    \includegraphics[width=\textwidth]{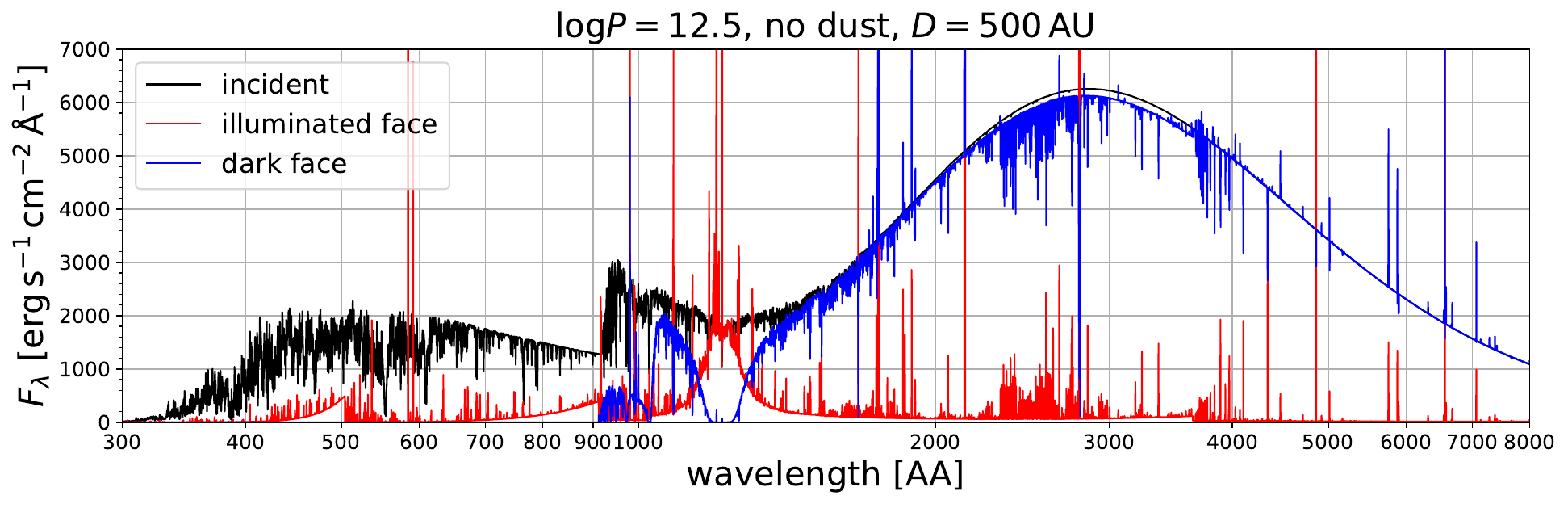}
    \caption{Spectra for the three \texttt{CLOUDY} models shown in \reffig{cloudy_models}. We show the incident stellar radiation (black), emission from the illuminated slab face (red), and transmitted radiation emerging from the rear face if one views the gas slab against the stellar source (blue). In the dust free models, the slab has a very small optical depth to Balmer continuum photons due to photo-ionization of the $n=2$ hydrogen atoms.}
    \label{fig:cloudy_models_spectra}
\end{figure*}

In \reffig{cloudy_models_spectra}, we plot the corresponding radiation spectra computed for these models. For the dust-free models, and up to the gas column that is computed, $\log N_{\rm H}=10^{22}$--$10^{22.5}$, the predominantly neutral gas slab is optically thin to Balmer continuum photons, which means only a small fraction of them contribute to $n=2$ photo-ionization and heating. Overall, the \texttt{CLOUDY} calculations indicate that the partially ionized zone can form in a gas rich in metals.

\section{Discussion}
\label{sec:discuss}

In this Section, we discuss several additional aspects of the partial ionization phenomenon.

\subsection{Balmer Line Ratio}
\label{sec:balmer_line_ratios}

The emergent spectrum of the partially ionized zone should show a H$\alpha$/H$\beta$ line ratio much larger than from ordinary H II regions. While we have shown that this zone is optically quite thin to Balmer continuum photons, the optical depths for Balmer line photons are typically large, for example in the range of $\mathcal{O}(10^2$-$10^3)$ for H$\alpha$. This differs from the Case C recombination defined in \cite{Xu1992HRecombinationSN1987A}, which has large Balmer continuum optical depths. This causes the $n=3$ atoms to be much more abundant than the $n=4,\,5,\cdots$ atoms, as can be seen from the one-zone calculations in \reffig{slab} and the \texttt{CLOUDY} calculations in \reffig{cloudy_models}. Consequently, the H$\beta$ emissivity is much smaller than that for H$\alpha$. This is unlike in the usual H II zone, where the gas is optically thin to Balmer line photons, and the similar order-of-magnitude values for the radiative rates of permitted lines lead to more comparable populations of $n=3,\,4,\,5,\,\dots$ following recombination cascades. We note that for $\log\nH=7$--$9$, collisions are not fast enough compared to spontaneous radiative transitions, even after accounting for line trapping, to drive state populations to thermal equilibrium values, especially for $n\geqslant 3$.

The elevation in the $n=3$ population relative to $n=4$ is clearly seen for the three \texttt{CLOUDY} models shown in \reffig{cloudy_models}. The three models, from the left to the right column, exhibit an emergent H$\alpha$/H$\beta$ line ratio that equals $4.5$, $9.8$ and $12.1$, respectively, with emissions from both the H II zone and the partially ionized zone combined. These ratios are much higher than the typical H$\alpha$/H$\beta$ ratio $\sim 2.7$--$3.0$ for Case B recombination in an H II region of low density and small Balmer line optical depths~\citep{Draine2011ISMtext}.

We note that the Weigelt blobs show a remarkably large observed H$\alpha$/H$\beta$ ratio $\sim 27$~\citep{Zethson2012Weigeltlines}, a factor of $\sim 9$ larger than what is expected from usual Case B recombination. This huge Balmer decrement is only partially due to foreground dust reddening. Analyzing optical [Fe II] and [Ni II] line ratios, \cite{Hamann1999HSTSTISspectraEtaCarEjecta} found $E_{B-V}=0.6$--$0.8\,$mag, which means $A_V=1.9$--$2.5\,$mag for $R_V=3.1$. Fluxes measured for a large number of optical [Fe II] lines suggest that $A_V<2.6\,$mag~\citep{Hamann2012EtaCarInnerEjecta}. \cite{Damineli2019EtaCar} estimated a total foreground $A_V=2.45\,$mag by separating ISM and intracluster reddening contributions \citep{Hur2012YoungOpenClustersCarinaNebua}. These $A_V$ values account for up to a factor of $\sim 3$--$4$ in the H$\alpha$/H$\beta$ ratio, which implies H$\alpha$/H$\beta$=$7$--$9$ intrinsically. Interestingly, this range is very consistent with our \texttt{CLOUDY} model for $\log P=12$ with $<10\%$ of the grains surviving (see \refsec{dust}).

Similarly, \cite{Choe2024} reported an H$\alpha$/H$\beta$ flux ratio $\approx 5$ for Godzilla, which was interpreted as significant dust reddening $E_{B-V}=0.45\,$mag (assuming $R_V=4.05$) in the local gaseous environment. Nevertheless, it is puzzling that the FUV continuum appears blue and FUV emission lines are strong. We suggest that this could instead be evidence for partially ionized gas. For a comparison, the nebula of the LyC-leaking super star cluster in the same Sunburst galaxy~\citep{RiveraThorsen2019Sci} has a lower density $\nele\approx 10^5\,{\rm cm}^{-3}$~\citep{Pascale2023SunburstLyCSSC}, and the observed H$\alpha$/H$\beta$ line ratio $\simeq 3$~\citep{RiveraThorsen2024SunburstLyCSSC} is indeed consistent with the usual Case B recombination in an H II zone. In general, Balmer line-ratio analyses for dust extinction diagnosis need to be performed with caution when the nebula is dense $\nele \gtrsim 10^6\,{\rm cm}^{-3}$.

\subsection{Effect of Internal Dust}
\label{sec:dust}

Dust grains can absorb Lyman series photons and reduce pumping of the excited states. In particular, they internally destroy Ly$\alpha$ photons and reduce the $n=2$ population. In the EPA, the effect of grains on Ly$\alpha$ photon transfer becomes important if $\varepsilon\,\tau_0 \gtrsim 1$ in \refeq{EPA_beta_eff}. The Ly$\alpha$ line center optical depth is
\begin{align}
    \tau_0 = 8 \times 10^8\,(1-\xe)\,\left(\frac{N_{\rm H}}{10^{22}\,{\rm cm}^{-2}}\right)\,\left(\frac{\Te}{6000\,{\rm K}}\right)^{-\frac12}.
\end{align}
The $\varepsilon$ parameter is
\begin{align}
    \varepsilon\,\tau_0 = 14\,f_{\rm d}\,\left(\frac{N_{\rm H}}{10^{22}\,{\rm cm}^{-2}}\right).
\end{align}
Hence if $f_{\rm d} < 0.07\,(N_{\rm H} / 10^{22}\,{\rm cm}^{-2})^{-1}$, destruction of Ly$\alpha$ photons by grains is unimportant for the formation of the partially ionized zone. While this requires a dust-to-gas ratio significantly lower than typical ISM values at solar metallicity, thorough removal of grains is not necessary.

\cite{Draine2011RadiationPressureHIIRegion} studied radiation pressure on dust grains in H II regions, but concluded that at $\nH \lesssim 10^3\,{\rm cm}^{-3}$ radiation does not push grains out of the H II zone fast enough compared to the lifetime of the ionizing stars $\sim 1\,$Myr. However, at much higher densities (also subject to much higher radiation intensity) grains can be pushed out of the H II zone or even drift through the H I gas over the nebula lifetime~\citep{PascaleDai2024godzilla}. It may be that grains have been removed from the partially ionized zone inside the Weigelt blobs by radiation over the $\sim 10^2\,$yr timescale.

The estimation of \cite{PascaleDai2024godzilla} is in a sense simplistic as grains are assumed to drift relative to a static gas. The Weigelt blobs probably have internal turbulence motion as commonly found for H II regions~\citep{GarciaVazpuez2023TurbulenceHIIregions}. The strong FUV luminosity of $\eta$ Car A implies a supersonic drift velocity
\begin{align}
    v_d = & 40\,{\rm km}\,{\rm s}^{-1}\,\left(\frac{L_{\rm FUV}}{5\times 10^6\,L_\odot}\right)^{\frac12}\,\left(\frac{\nH}{10^8\,{\rm cm}^{-3}}\right)^{-\frac12}\nonumber\\
    & \times \left(\frac{D}{500\,{\rm AU}}\right)^{-1},
\end{align}
which is likely faster than internal turbulence. We caution that there is much uncertainty on how efficiently grains can truly be removed by radiation with internal turbulence dragging them around randomly. Alternatively, other dust destruction mechanisms may be at play, such as radiative torques~\citep{Hoang2019NatAsRadiativeTorqueGrainDestruction}.

Another effect of dust under UV irradiation is photoelectric heating. We follow \cite{WeingartnerDraine2001GrainChargingPhotoheating} to assess the importance. \cite{WeingartnerDraine2001GrainChargingPhotoheating} defines the quantity $G\,\sqrt{\Te}/\nele = (u_{\rm UV}/u_0)\,\sqrt{\Te}/\nele$ ($G$ is not Newton's gravitational constant), where $u_{\rm UV}$ is the local UV radiation energy density and $u_0=5.33\times 10^{-14}\,{\rm erg}\,{\rm cm}^{-3}$ is a reference value. For the Weigelt blobs, we estimate
\begin{align}
    & \frac{G\,\sqrt{\Te}}{\nele} = 1.2 \times 10^5\,{\rm K}^{1/2}\,{\rm cm}^3\,\left(\frac{L_{\rm UV}}{5 \times 10^6\,L_\odot}\right)\,\left(\frac{D}{500\,{\rm AU}}\right)^{-2}\nonumber\\
    & \times \left(\frac{\Te}{5000\,{\rm K}}\right)^{\frac12}\,\left(\frac{\nele}{10^7\,{\rm cm}^{-3}}\right)^{-1}.
\end{align}
For the above high $G\,\sqrt{\Te}/\nele$ value, Figure 12 and 13 in \cite{WeingartnerDraine2001GrainChargingPhotoheating} imply that about $10^{-4}$--$10^{-3}$ of the UV radiation absorbed by the grains contributes to net gas heating. The relatively low efficiency is due to quenching of photo-electron emission when the grain is charged to a high potential. For a dusty gas, this may be a significant or even dominant contribution to gas heating behind the H II zone, thus helping with the formation of partial ionization. UV radiation can only penetrate a limited depth into a dusty gas, so the effect of Raman scattering will be limited. However, reliable predictions are complicated by the radiation pressure effects acting on grains.

\subsection{Other Heating Processes}
\label{sec:otherheating}

If additional gas heating mechanisms operate at low $\Te$, a stable equilibrium of lukewarm, partially ionized gas may exist even without the effect of Ly$\alpha$ injection via Raman scattering. Below we discuss possible processes in the context of the Weigelt blobs.

First of all, it is known \citep[e.g.][]{Ferland2003SpectrumOfSingleCloud} that AGN BLRs acquire partially ionization because copious X-ray photons from the AGN central source penetrate beyond the usual H ionization front. They keep the gas warm and maintain the population of excited H atoms required for photo-ionization by UV photons. With such energy source, a net supply of Ly$\alpha$ photons is provided through collisional excitation of $n=2$ atoms, even without contributions from the Raman processes. Such strong X-ray irradiation is absent for the Weigelt blobs and is unlikely relevant for nebulae powered by star formation.

The warm interior of the Weigelt blobs may have strong internal turbulence, which may be supersonic and can dissipate kinetic energy. The volumetric heating rate can be roughly estimated as
\begin{align}
    \Gamma_{\rm tb} \sim & m_p\,\nH\,\frac{v^3_{\rm tb}}{l} = 7 \times 10^{-12}\,{\rm erg}\,{\rm s}^{-1}\,{\rm cm}^{-2} \nonumber\\
    & \times \left(\frac{\nH}{10^8\,{\rm cm}^{-3}}\right)\,\left(\frac{v_{\rm tb}}{40\,{\rm km}\,{\rm s}^{-1}}\right)^3\,\left(\frac{l}{100\,{\rm AU}}\right)^{-1}.
\end{align}
where the largest eddies have a size $l$ and a fluid velocity $v_{\rm tb}$. Applying this to models in \reffig{slab}, turbulence heating is small compared to radiative contributions, even for unrealistic turbulence velocity values $v_{\rm tb}=40\,{\rm km}\,{\rm s}^{-1}$ that is already comparable to the bulk velocity of the blobs~\citep{Hamann2012EtaCarInnerEjecta}.

Cosmic ray (CR) ionization can contribute to gas heating. The net heating rate due to CRs interacting with H I is~\citep{DalgarnoMcCray1972ARAAHeatingIonizationHI}
\begin{align}
    \Gamma_{{\rm CR},n} \approx & 1 \times 10^{-20}\,{\rm erg}\,{\rm s}^{-1}\,\left(\frac{\nH}{10^7\,{\rm cm}^{-3}}\right)\,\left(\frac{\zeta_{\rm CR}}{10^{-16}\,{\rm s}^{-1}}\right)\nonumber\\
    & \times \left[1 + 4.06\,\left(\frac{\xe}{\xe+0.07}\right)^{\frac12}\right].
\end{align}
Here the fiducial value for the CR primary ionization rate $\zeta_{\rm CR}$ is typical in average Galactic ISM. However, it is plausible that in a highly dynamic nebula environment, such as $\eta$ Car's inner ejecta, the local CR density may be orders of magnitude higher~\citep{Steinmassl2023CosmicRayEscapeFromEtaCar}.

For the examples in \reffig{slab}, CR heating becomes comparable to the contribution from Raman scattering, for low $\Te$, if locally $\zeta_{\rm CR} \sim 10^{-7}\,{\rm s}^{-1}$. In that case, CR heating by itself will allow a warm equilibrium. However, such high $\zeta_{\rm CR}$ translates into a huge CR energy flux $\sim 10^6\,{\rm erg}\,{\rm s}^{-1}\,{\rm cm}^{-2}$. By comparison, the primary star drives an intense wind with a mass loss rate up to $\dot M_{w, {\rm A}} \approx 10^{-3}\,M_\odot\,{\rm yr}^{-1}$ at a terminal velocity $v_{w, {\rm A}} \approx 600\,{\rm km}\,{\rm s}^{-1}$~\citep{Hillier2001EtaCar, Groh2012EtaCar2DWindCollisionModel, Clementel2015EtaCar3DWindCollisionModel}; the wind of the secondary star is much more dilute $\dot M_{w, {\rm B}} \approx 10^{-5}\,M_\odot\,{\rm yr}^{-1}$ and $v_{w, {\rm B}} \approx 3000\,{\rm km}\,{\rm s}^{-1}$~\citep{PittardCorcoran2002EtaCarWindParameters, Parkin2009EtaCarCollidingWinds3DModel}, and has a subdominant mechanical power~\citep{Grant2023EtaCarWinds}. This provides a total kinetic energy flux $\sim 1.6 \times 10^5\,{\rm erg}\,{\rm s}^{-1}\,{\rm cm}^{-2}\,(D/500\,{\rm AU})^{-2}$, only a small fraction of which is expected to accelerate CRs at shocks (radiation energy flux from the primary star is much higher $\sim 2.7 \times 10^7\,{\rm erg}\,{\rm s}^{-1}\,{\rm cm}^{-2}\,(D/500\,{\rm AU})^{-2}$). Therefore, it is unlikely that the Weigelt blobs are exposed to a CR background sufficiently intense to be the dominant energy source to maintain a warm zone of partial ionization.

In nebulae powered by non-eruptive massive stars, it is typical that the total mechanical luminosity of the system (e.g. winds and outflows) amounts to a small fraction of the UV luminosity. Hence, if CRs are accelerated locally, from the consideration of energetics we can expect that CR heating of the H I gas is unlikely to be important compared to radiative heating mechanisms including photo-ionization and Raman scattering.

\section{Conclusion}
\label{sec:concl}

We have studied the formation of a partially ionized zone behind the usual H II zone, in a gas slab photoionized by incident continuum radiation. Such a partially ionized zone is conducive to the internal formation of a scattering-broadened Ly$\alpha$ line, which can power strong fluorescent emissions from metals via the Bowen mechanism as observed from many dense nebulae.

We have developed a simple one-zone model of hydrogen gas to unveil the underlying atom-radiation interaction (\refsec{one_zone}). Lyman continuum photons are exhausted and do not contribute to ionization. Instead, ionization is primarily sustained by incident Balmer continuum photons ionizing $n=2$ hydrogen atoms, which are pumped by trapped Ly$\alpha$ photons.

To maintain a steady state, Ly$\alpha$ photons must be replenished. Without photoionization from the ground state, recombination does not replenish Ly$\alpha$ photons. Electron collisional excitations of $n=2$ can do the job, which however requires a high $\Te$ at which photo-heating falls short of radiative cooling unless very high density $\nH \gtrsim 10^9\,{\rm cm}^{-3}$ suppresses cooling sufficiently. 

We point out that for UV-intense radiation spectra, Raman scattering of incident FUV photons on the damping wings of Ly$\beta$, Ly$\gamma$, Ly$\delta$, $\cdots$ is a major mechanism to supply Ly$\alpha$ photons independent of the gas temperature (\refsec{Raman}). By deriving analytical results in the low temperature limit, we have explained that when the radiation-to-gas ratio is fixed, the ionization fraction increases roughly as $\nH^{-1/2}$ (\refsec{lowTe}). If the radiation-to-gas ratio is limited by the dynamic effect of LyC and UV photons pressing on the gas slab, significant ionization fraction $\xe \gtrsim 0.1$ only occurs for $\nH \gtrsim 10^7\,{\rm cm}^{-3}$. 

The partially ionized gas is often optically thin to Balmer continuum photons but thick to Balmer line photons. This leads to an H$\alpha$/H$\beta$ line ratio much higher than from ordinary H II regions, and has implications for inferring dust extinction from Balmer decrement in high density nebulae (\refsec{balmer_line_ratios}).

For the irradiation condition of the Weigelt blobs at $500\,$AU from the $\eta$ Car system and a range of gas density $\nH=10^7$--$10^9\,{\rm cm}^{-3}$, our simple one-zone hydrogen gas calculations indicate that partial ionization arises at $\Te \approx 6000$K when Raman scattering is accounted for, falling on the lower end of the range $\Te=6000$--$8000\,$K inferred from observation. If Raman scattering is absent, the blobs would have to be either denser than $10^9\,{\rm cm}^{-3}$ or much closer to the central binary. This is qualitatively confirmed by 1D \texttt{CLOUDY} calculations that include realistic chemical compositions and stratified temperature and ionization structures for the Weigelt blobs, although a lower $\Te=5000\,$K is found in these calculations.

In Galactic stellar nurseries, proplyds are dense gas disks around young stars photo-ionized by neighboring massive stars. Some proplyds in the Orion Nebula host dense ionized gas with $\nele \sim 10^6\,{\rm cm}^{-3}$~\citep{Henney2002LV2, Hodapp2009OrionisIRS1ABproplyd}, with the shielded H I interior probably reaching $\nH \gtrsim 10^7\,{\rm cm}^{-3}$. It is interesting to see with UV telescopes whether dense proplyds exhibit Ly$\alpha$-pumped fluorescent emission lines, as a test of the partial ionization mechanism studied here.

Looking into the more distant universe, this phenomenon might also take place in the nebula powered by the Cosmic Noon super star cluster candidate ``Godzilla'', for which Ly$\alpha$ pumped Fe fluorescent lines appear remarkably strong~\citep{Vanzella2020Tr, Choe2024} and a very high electron density $\nele \approx 10^{7.5}\,{\rm cm}^{-3}$ has been inferred from photo-ionization modeling~\citep{PascaleDai2024godzilla} and is responsible for the observed unusual collisional line ratios~\citep{PascaleDai2024godzilla, Choe2024}. Another super star cluster in the same galaxy hosts a dense, nitrogen-enriched nebula in its proximity~\citep{Pascale2023SunburstLyCSSC, RiveraThorsen2024SunburstLyCSSC}, although the inferred $\nele \sim 10^5\,{\rm cm}^{-3}$ should be too low to allow partial ionization. A local analog is the nebula powered by the star-forming core of the Blue Compact Dwarf galaxy Mrk 996~\citep{Thuan1996Mrk996HST, Telles2014Mrk996GMOS}. A high density $\nele\sim 10^{7}\,{\rm cm}^{-3}$ is inferred from optical spectroscopy~\citep{James2009Mrk996VIMOS}. On the other hand, some high-$z$ galaxies have nebular gas at unusually high densities. One example is the galaxy GN-z11 at $z=10.6$~\citep{Bunker2023GNz11JADESNIRSpec, Senchyna2024GNz11Nitrogen}, with evidence for $\nele \sim 10^9\,{\rm cm}^{-3}$ and a possible origin from the AGN Broad Line Region~\citep{Maiolino2024GNz11AGN}. \cite{Topping2024CIVandNEmittersReionizationEra} reported another galaxy A1703-zd6 whose ISM approaches $\nele \sim 10^5\,{\rm cm}^{-3}$. It is intriguing that all these dense nebula exhibit enhancement in the nitrogen abundance~\citep{MarquesChaves2024HighRedshiftNitrogenEmitters}. The underlying physical connection between high density and self-enrichment remains to be understood. Partially ionized zones may form in those nebulae as well, which is an interesting question for future study. Possible observational signatures include Bowen fluorescent emission powered by Ly$\alpha$, and abnormally large H$\alpha$/H$\beta$ ratios that are inconsistent with other dust reddening indicators.

Recently, \cite{InayoshiMaiolino2024BalmerBreakAGNBLR} ran \texttt{CLOUDY} to study photo-ionization at high gas densities typical of AGN Broad Line Regions (BLRs) in the context of understanding the spectra of the Little Red Dots discovered by JWST at high $z$. The authors find that when $\nH\gtrsim 10^9\,{\rm cm}^{-3}$ the hydrogen $n=2$ population is high enough behind the H ionization front that a clear Balmer break is expected to show in the transmission spectrum of the continuum source. In their models with $\nH< 10^9\,{\rm cm}^{-3}$, the Balmer break is not conspicuous, even though the $n=2$ population is still significant. This work focuses on densities still several orders of magnitude lower than that of AGN BLR clouds, for which the gas slab typically has a small optical depth to Balmer continuum photons.

\section*{acknowledgments}

The author is grateful to William Henney for carefully reading the draft version of this work and providing constructive comments. The author would also like to thank Gary Ferland, Christopher Matzner, Christopher McKee, and Massimo Pascale for useful discussions. L.D. acknowledges research grant support from the Alfred P. Sloan Foundation (Award Number FG-2021-16495), and support of Frank and Karen Dabby STEM Fund in the Society of Hellman Fellows.

\appendix

\section{Statistical Equilibrium}
\label{app:equil}

Let $\xe=n_{\rm HII}/\nH = \nele/\nH$ be the ionization fraction of the hydrogen gas, $x_n = n_{\rm HI, n}/\nHI$ be the fractional population of the $n$-th (ground or excited) states. Explicitly including a finite tower of states $n=1,\,2,\,3,\,\cdots,\,n_{\rm max}$ in our hydrogen atom model, there is the constraint $\sum^{n_{\rm max}}_{n=1}\,x_n = 1$.

In statistical equilibrium, we have the following set of linear equations for $x_n$'s:
\begin{eqnarray}
\label{eq:eq_equil_pop}
    0 & = & (1-\xe)\,\xe\,\nH\,\left[ \sum^{n-1}_{n'=1}\,q_{n,n'}(\Te)\,\left(x_n - x_{n'}\frac{g_n}{g_{n'}}\,e^{-\frac{E_{n,n'}}{\kB\,\Te}}\right) - \sum^{n_{\rm max}}_{n'=n+1}\,q_{n',n}(\Te)\,\left(x_{n'} - x_{n}\frac{g_{n'}}{g_{n}}\,e^{-\frac{E_{n',n}}{\kB\,\Te}}\right) \right] \nonumber\\
    && + (1 - \xe)\,\left[ x_n\,\sum^{n-1}_{n'=1}\,A_{n,n'}\,\beta^{\rm eff}_{n,n'} - \sum^{n_{\rm max}}_{n'=n+1}\,x_{n'}\,A_{n',n}\,\beta^{\rm eff}_{n',n}\right]  - \xe^2\,\nH\,\alpha^{\rm rec}_{n}(\Te) + (1-\xe)\,\xe\,x_n\,\nH\,q^{\rm ci}_n(\Te) \nonumber\\
    && + (1-\xe)\,x_n\,\int\rmd\nu\,\frac{F_\nu(\nu)}{h\nu}\,\sigma^{\rm pi}_n(\nu) + (1-\xe)\,\frac{1}{4}\,\delta_{n2}\,x_n\,A_{2s} - (1 - \xe)\,x_n\,A^{\rm Ra}_n.
\end{eqnarray}
Here $g_n=2 n^2$ is the statistical weight, and $E_{n, n'} = E_n - E_{n'}$ (for $n>n'$) is the energy difference between the $n$-th state and the $n'$-th state. 

The first line describes (de-)excitation by electron collision, where $q_{n,n'}(\Te)$ is the rate coefficient for collisional de-excitation $n\rightarrow n'$. 

In the second line, the first two terms account for spontaneous and stimulated radiative transitions between states, where $A_{n,n'}$ is the ($l$-level averaged) Einstein coefficient for spontaneous radiative decay $n\rightarrow n'$. In the Escape Probability Approximation (EPA), the sum of spontaneous emission and stimulated emission and absorption of a resonant line is accounted for by the effective escape probability $\beta^{\rm eff}_{n,n'}$, for transition $n\rightarrow n'$, which depends on the line optical depth and dust absorption (\refeq{EPA_beta_eff}). The third term accounts for radiative recombination, with $\alpha^{\rm rec}_n(\Te)$ being the partial rate coefficient for recombination into the $n$-th state. Since the gas slab is extremely optically thick to Lyman continuum photons, we will set $\alpha^{\rm rec}_1(\Te)=0$ as in Case B. The fourth term accounts for ionization by electron collision, with $q^{\rm ci}_n(\Te)$ being the rate coefficient from the $n$-th state.

In the third line, the first term describes photoionization by external continuum radiation $F_\nu(\nu)$, where $\nu$ is the photon frequency and $\sigma^{\rm pi}_n(\nu)$ is the frequency-dependent photoionization cross section for the $n$-th state. Since we consider a situation in which Lyman continuum photons are all consumed by an H II zone in the front, there will be no photoionization from the ground state. The second term accounts for two-photon decay of the $2s$ state, which is assumed to have $1/4$ of the $n=2$ population. The third term, proportional to the injection rate $A^{\rm Ra}_n$ [${\rm s}^{-1}$] into the $n$-th state, accounts for injection contribution from Raman scattering of external continuum photons on the damping wings of higher-order Lyman series lines.

For given $\xe$ and $\Te$, \refeq{eq_equil_pop} can be solved to find the $x_n$'s, for $n=1,2,\cdots,n_{\rm max}$. To find the equilibrium value of $\xe$, we need to consider the balance between ionization and recombination. An additional equation is required to determine $\xe$,
\begin{align}
    (1-\xe)\,\int\rmd\nu\,\frac{F_\nu(\nu)}{h\nu}\,\sum^{n_{\rm max}}_{n=1}\,\sigma^{\rm pi}_n(\nu)\,x_n + (1-\xe)\,\xe\,\nH\,\sum^{n_{\rm max}}_{n=1}\,x_n\,q^{\rm ci}_n(\Te) = \xe^2\,\nH\,\sum^{n_{\rm max}}_{n=2}\,\alpha^{\rm rec}_n(\Te).
\end{align}
As external continuum radiation dominates the photoionization rate, we neglect photoionization by photons produced inside the gas slab (via radiative recombination or radiative transition between states).

\section{Energy Balance}
\label{app:energy}

The equilibrium gas temperature $\Te$ must be determined from the balance of gas heating and cooling, which can be calculated by accounting for contributions from photo-ionization, free-bound and free-free emissions, and collisional excitation of resonant lines. At statistical equilibrium, a different but equivalent approach to find balance between heating and cooling is to equate the rate of energy injection into the gas slab with the rate of energy loss, both of which are due to photons. We will pursue this second approach.

The rate of energy injection is the sum of two primary contributions
\begin{align}
    \Gamma_{\rm tot} = \Gamma_{\rm pi} + \Gamma_{\rm Ra}
\end{align}
Photo-ionization injects energy into the gas at a rate
\begin{align}
    \Gamma_{\rm pi} = (1-\xe)\,\nH\,\int\rmd\nu\,F_\nu(\nu)\,\sum^{n_{\rm max}}_{n=1}\,\sigma^{\rm pi}_n(\nu)\,x_n.
\end{align}
For Raman scattering of external continuum photons, the energy injection rate is
\begin{align}
\label{eq:GaRa}
    \Gamma_{\rm Ra} =  (1 - \xe)\,\nH\,x_1\,\sum^{n_{\rm max}}_{n=2}\,A^{\rm Ra}_n\,E_{n,1}.
\end{align}

The rate of energy loss has several primary contributions
\begin{align}
    \Lambda_{\rm tot} = \Lambda_{\rm ff} + \Lambda_{\rm rr} + \Lambda_{2\gamma} + \Lambda_{\rm line}.
\end{align}
For free-free emission, we use a fitting formula from \cite{Draine2011ISMtext}
\begin{align}
    \Lambda_{\rm ff} = \xe^2\,\nH^2\,\alpha^{\rm rec}_B(\Te)\,\left( 0.54\,\kB\,\Te \right)\,\left(\frac{\Te}{10^4\,{\rm K}}\right)^{0.37}.
\end{align}
For radiative recombination, we approximate the rate of energy loss as
\begin{align}
    \Lambda_{\rm rr} = \xe^2\,\nH^2\,\sum^{n_{\rm max}}_{n=2}\,\alpha^{\rm rec}_n(\Te)\,\left(E^B_{\rm rr}(\Te) + \frac{I_H}{n^2} \right).
\end{align}
where $I_H = 13.59\,$eV is the ionization potential of hydrogen. We choose to use the standard Case B mean energy of recombining electrons~\citep{Draine2011ISMtext}
\begin{align}
    E^B_{\rm rr}(\Te) = \left[ 0.684 - 0.0416\,\ln\left(\frac{\Te}{10^4\,{\rm K}}\right) \right]\,\kB\,\Te.
\end{align}
The rate of energy loss due to two-photon decay of $2s$ is simply
\begin{align}
    \Lambda_{2\gamma} = \frac{3}{16}\,(1-\xe)\,\nH\,x_2\,A_{2s}\,I_H.
\end{align}
The last contribution is loss through resonant lines, either because the line photons directly escape the gas slab, or because they are absorbed by dust grains and re-emitted as infrared photons and then escape the slab. Within the EPA, this energy loss rate can be computed as
\begin{align}
    \Lambda_{\rm line} = (1-\xe)\,\nH\,\sum^{n_{\rm max}}_{n'=1}\,\sum^{n_{\rm max}}_{n=n'+1}\,x_n\,A_{n,n'}\,\beta^{\rm eff}_{n,n'}\,E_{n,n'}.
\end{align}

\section{Rayleigh and Raman scattering cross sections}
\label{app:RayleighRaman}

Analytic formulae for the cross sections of Rayleigh and Raman scattering for the hydrogen atom can be found in many references. A detailed discussion is recently presented in \cite{Kokubo2024RayleighRaman} (also see references therein). For completeness, we present here analytic expressions used in our calculations, which are avearged over the $l$-levels, account for hydrogen excited states up to $n=n_{\rm max}$, and correspond to the small hydrogen atom model we use. These expressions correspond to what are presented in \cite{Nussbaumer1989RamanScattering}, but without including the small correction from intermediate continuum states.

First, Rayleigh scattering off the hydrogen $n=1$ state has a cross section
\begin{align}
    \sigma_{\rm Rayl}(\nu) = \sigma_T\,\left|\sum^{n_{\rm max}}_{n=2}\mathcal{M}^{(n)}(\nu)\right|^2.
\end{align}
Here $\sigma_T=6.65 \times 10^{-25}\,{\rm cm}^2$ is the Thomson scattering cross section. The partial amplitude is given by
\begin{align}
    \mathcal{M}^{(n)}(\nu) = \frac{f_{n,1}}{E_{n,1}/(h\nu)-1},
\end{align}
where $f_{n,n'}$ stands for the ($l$-level averaged) oscillator strength for the transition $n\rightarrow n'$ ($n > n'$):
\begin{align}
    f_{n,n'} = A_{n,n'}\,\frac{g_n}{g_{n'}}\,\left(\frac{h\,c}{E_{n,n'}}\right)^2\,\frac{m_e\,c}{8\pi^2\,e^2}.
\end{align}

For the Raman scattering from the ground state to the $n$-th excited state ($n>1$), the cross section is give by
\begin{align}
    \sigma^{(n)}_{\rm Ra}(\nu) = \frac{1}{16}\,\sigma_T\,h\nu\,(h\nu - E_{n,1})^3\,\left|\sum^{n_{\rm max}}_{n'=n+1}\mathcal{M}^{(n',\,n)}_{\rm Ra}(\nu)\right|^2.
\end{align}
Here the partial amplitude corresponding to the final state principal quantum number $n$ and the virtual state quantum number $n'$ is given by
\begin{align}
    \mathcal{M}^{(n',\,n)}_{\rm Ra}(\nu) = \left(\frac{g_1\,f_{n',1}}{E_{n',1}}\right)^{\frac12}\,\left(\frac{\tilde g_n\,f_{n',n}}{E_{n',n}}\right)^{\frac12}\,\frac{(E_{n',1} + E_{n',n})}{(E_{n',n} + h\nu)\,(E_{n',1} - h\nu)}.
\end{align}
Here $g_1=2$ is the statistical weight for the ground state. The statistical weight for the final state requires care. For $n=2$, we use $\tilde g_2=2$ since the Raman final state can only be $2s$. For $n=3,\,4,\,\cdots$, we use $\tilde g_n=12$ because the Raman final state can be either $ns$ or $nd$.

\section{Low temperature regime}
\label{app:LowTe}

If $\Te$ is very low, couplings between the different $n$ states of hydrogen atom as well as ionization by electron impact all become inefficient. It is possible to derive analytic approximations for the one-zone hydrogen slab in this regime.

We shall assume that nearly all hydrogen atoms are in the $n=1$ ground state, $x_1\approx 1$. Assuming that photo-ionization is primarily from $n=2$, the balance between radiative recombination and photo-ionization requires
\begin{align}
    \xe^2\,\nH^2\,\alpha^{\rm rec}_B(T) = x_2\,(1-\xe)\,\nH\,\int\,\rmd\nu\,\frac{F_\nu(\nu)}{h\nu}\,\sigma^{\rm pi}_2(\nu).
\end{align}
Each radiative recombination event cascades to $n=2$ under the Case B condition, so that the steady state for the $n=2$ population requires
\begin{align}
    0 = & - x_2\,(1-\xe)\,\nH\,A_{2,1}\,\beta^{\rm eff}_{2,1} - \frac14\,x_2\,(1-\xe)\,\nH\,A_{2s} - x_2\,(1-\xe)\,\nH\,\int\,\rmd\nu\,\frac{F_\nu(\nu)}{h\nu}\,\sigma^{\rm pi}_2(\nu) \nonumber\\
    & + \xe^2\,\nH^2\,\alpha^{\rm rec}_B(\Te) + \frac{1}{d}\,\int\rmd\nu\,\frac{F_\nu(\nu)}{h\nu}\,f_{\rm Ra}(\nu),
\end{align}
where the last term accounts for excitation to $n=2$ via Raman scattering. From these two equations, we find
\begin{align}
    \xe^2 = \frac{1}{\nH^2\,\alpha^{\rm rec}_B(\Te)\,d\,\left(A_{2,1}\,\beta^{\rm eff}_{2,1}+A_{2s}\right)}\,\left(\int\,\rmd\nu\,\frac{F_\nu(\nu)}{h\nu}\,\sigma^{\rm pi}_2(\nu)\right)\,\left(\int\rmd\nu\,\frac{F_\nu(\nu)}{h\nu}\,f_{\rm Ra}(\nu)\right).
\end{align}
It may seem like the right hand side could grow larger than unity if either incident radiation is arbitrarily increased or if the slab thickness $d$ is arbitrarily reduced; in fact, when $\xe$ approaches unity, the Raman rate will be limited since the H I column is proportional to $d\,(1-\xe)$ and would become very small. This means that in that limit, $f_{\rm Ra}(\nu)$ decreases in proportional to $d\,(1-\xe)$. In light of this, we can write
\begin{align}
    f_{\rm Ra}(\nu) = (1-\xe)\,d\,\nH\,\sigma_{\rm Ra}(\nu)\,\frac{f_{\rm Ra}(\nu)}{\tau_{\rm Ra}(\nu)},
\end{align}
where $\sigma_{\rm Ra}(\nu)$ is the total Raman scattering cross section from $1s$, and the ratio $f_{\rm Ra}(\nu)/\tau_{\rm Ra}(\nu)$ is close to unity if the slab is optically thin to Raman scattering $\tau_{\rm Ra}(\nu) \ll 1$, and decreases to zero as $1/\tau_{\rm Ra}(\nu)$ if the slab is optically very thick to Raman scattering $\tau_{\rm Ra}(\nu) \gg 1$. Then we obtain
\begin{align}
    \frac{\xe^2}{1-\xe} = \frac{1}{\nH\,\alpha^{\rm rec}_B(\Te)\,\left(A_{2,1}\,\beta^{\rm eff}_{2,1}+A_{2s}\right)}\,\left(\int\,\rmd\nu\,\frac{F_\nu(\nu)}{h\nu}\,\sigma^{\rm pi}_2(\nu)\right)\,\left(\int\rmd\nu\,\frac{F_\nu(\nu)}{h\nu}\,\sigma_{\rm Ra}(\nu)\,\frac{f_{\rm Ra}(\nu)}{\tau_{\rm Ra}(\nu)}\right),
\end{align}
which is \refeq{xe_lowTe}.

\bibliography{refs}{}
\bibliographystyle{aasjournal}

\end{document}